\documentclass[a4paper,11pt]{article}
\usepackage[latin1]{inputenc}
\usepackage{indentfirst}

\usepackage{amsfonts}
\usepackage{float}
\usepackage{latexsym,amssymb,amsmath}
\usepackage[dvips]{graphics,graphicx,epsfig,color}
\usepackage[lofdepth,lotdepth]{subfig}
\begin{document}
\thispagestyle{empty}
\title{\bf Effective medium description of the resonant elastic-wave response at the periodically-uneven boundary of a half-space }
\author{Armand Wirgin\thanks{LMA, CNRS, UMR 7031, Aix-Marseille Univ, Centrale Marseille, F-13453 Marseille Cedex 13, France, ({\tt wirgin@lma.cnrs-mrs.fr})} }
\date{\today}
\maketitle
\begin{abstract}
A periodically-uneven (in one horizontal direction) stress-free boundary covering a linear, isotropic, homogeneous, lossless solid half space is submitted to a vertically-propagating shear-horizontal plane, body wave. The rigorous theory of this elastodynamic scattering problem is given and the means by which it can be numerically solved are outlined. At quasi-static frequencies, the solution is obtained from one linear equation in one unknown. At higher, although still low, frequencies, a suitable approximation of the solution is obtained from a system of two linear equations in two unknowns. This solution is shown to be equivalent to that of the problem of a vertically-propagating shear-horizontal plane body wave traveling in the same solid medium as before, but with a linear, homogeneous, isotropic layer replacing the previous uneven boundary. The thickness of this layer is equal to the vertical distance between the extrema of the boundary uneveness and the effective body wave velocity therein is equal to that of the underlying solid, but the effective shear modulus of the layer, whose expression is given in explicit algebraic form, is different from that of the underlying solid, notably by the fact that it is dispersive and lossy. It is shown that this dispersive, lossy effective layer, overriding the nondispersive, lossless solid half space, gives rise to two distinctive features of low-frequency response: a Love mode resonance and a Fixed-base shear wall pseudo-resonance. This model of effective layer with dispersive, lossy
 properties, enables simple explanations of how the low-frequency resonance and pseudo-resonance vary with  the geometric parameters (and over a wide range of the latter) of the uneven boundary.
\end{abstract}
Keywords: dynamic response, effective medium, boundary uneveness.
\newline
\newline
Abbreviated title: An effective layer that responds like a grating to a plane wave
\newline
\newline
Corresponding author: Armand Wirgin, \\ e-mail: wirgin@lma.cnrs-mrs.fr
 \newpage
\tableofcontents
\newpage
\section{Introduction}\label{intro}
This study originated in the search (ongoing since the last sixty years) for simple explanations (largely-lacking until now) of the main features (amplitude and spectral properties)   of seismic response (vibrations at locations near and/or above the ground) in above-sea level natural  geophysical configurations such as individual, or groups of, ridges, hills, mountains, etc.
\cite{al70,ba82,bc09,brt99,bo73,bo72,bp05,be14,bf14,ejp57,gb88,gb79,ke01,kb91,lk09,ma07,mca14,plb94,pi96,ro06,ssc91,si78,tf15,wi88,wi89,wkb92,wkb93}.
Most of these studies have been either of experimental or numerical nature. The experimental studies usually concerned in situ configurations, so that  considering the extreme diversity of the latter, there is little hope of deriving general laws from them. The numerical studies usually deal with periodic, small-height interface or boundary uneveness idealizations of the geophysical configurations (as well as of other elastic-wave devices \cite{ma88,tj03,wi88}), and  have not, either, given rise  to satisfactory explanations of many (including low-frequency) features of the response.

Since earthquakes also affect man-made, above-ground, structures such as buildings, city blocks and even whole cities, which are, in a sense similar to small-scale hills or mountains, there has been some research  on this question too \cite{br04,gr05,gtw05,gw08,ks06,ro06,tr72,wi16,wi18a,wi18c,wb96}. Contrary to the studies concerning the natural uneven boundary  configurations, some publications (e.g., \cite{br04,ro06}), based on homogenization techniques resulting in the reduction of the uneven boundary to a flat boundary loaded by a periodic set of mass-spring oscillators (analogous to fixed-base shear walls (FBSW) \cite{tr72}), have enabled to account for some of the main features (including what appear to be FBSW resonances)  of long wavelength (with respect to the characteristic dimensions of the representative features as well as their separation) seismic response of the uneven boundary representing a city.

Homogenization of empirical nature is a very old and even contemporary practice in geophysical problems. At present, homogenization has developed into a full branch of applied mathematics \cite{blp78,fb05,go02,kpv81,kr05,nk97} which has attracted the attention of physicists (mostly solid-state) interested in developing materials with unusual properties \cite{cs13,fb05,qk11,si02,vt17,wi18b}. These so-called 'metamaterials' are usually composed of periodic, or nearly- periodic assemblies of resonating elements similar to the blocks or buildings (thought of as single-degree-of-freedom oscillators) in \cite{br04}. A  feature of homogenization, which is very useful as a predictive tool, is that it relies on (e.g., see the discussion of the NRW technique in \cite{qk11}, or results in  \cite{si02,vt17})  the equivalence, as regards  low-frequency response to a wave, of the original geometrically-complex (although usually periodic) medium, boundary or interface  to  a geometrically-simpler, homogeneous medium or flat boundary. This is obtained at the expense of rendering more complex the constitutive properties of the media that are involved, but this complexity is precisely what accounts for the unusual (e.g., anomalous dispersion) response of the configuration.

Herein, we focus our attention on a very simple model of an uneven boundary (e.g., a mountain range, with periodicity along one horizontal coordiante) separating air from a solid underlying medium) submitted to a vertically propagating seismic wave. We show, by an effective medium  method that is somewhat similar to the NRW technique \cite{qk11}, that this uneven boundary responds, as concerns its amplitude and spectral features at low (but beyond static) frequencies, in much the same manner as a homogeneous layer whose upper and lower horizontal faces correspond to the highest and lowest  planes of the uneven boundary. In particular, we show  theoretically, and verify numerically that the low-frequency resonant response of the uneven boundary is dominated by the excitation of what is similar to a Love mode and a fixed-base shear-wall mode. Finally, we give mathematically-explicit expressions for the constitutive parameters of the effective layer.
\section{Exact solution of the problem of the response of the uneven boundary to the plane-wave solicitation}\label{exact}
%
\subsection{The boundary-value problem}\label{bvp}
In a cartesian coordinate system $Oxyz$, with origin at $O$, the uneven, on the average flat and  horizontal,  boundary  $\mathcal{B}$ separates two half-spaces, the upper  one $\mathcal{U}$ being occupied by the vacuum and the lower  one $\mathcal{L}$ being occupied by a homogeneous, isotropic, linear elastic solid. The uneveness of $\mathcal{B}$ is periodic in terms of $x$, oscillating between $z=0$ and $z=h$ with period $d$, and does not depend on $y$.

The elastic wave sources are assumed to be located in  $\mathcal{L}$  and to be infinitely-distant from $\mathcal{B}$ so that the  solicitation takes the form of a body (plane) wave in the neighborhood of $\mathcal{B}$. Its polarization is shear-horizontal ($SH$) so that only one (i.e., the $y$-) component of the incident displacement field is non-nil, i.e., $\mathbf{u}^{i}=(u_{x}^{i},u_{y}^{i},u_{z}^{i})=(0,u^{i}(\mathbf{x},\omega),0)$, wherein $\mathbf{x}=(x,z)$ and $\omega=2\pi f$  the angular frequency, $f$ the frequency.

Since neither the incident wavefield nor the geometric and compositional features of the configuration  depend on $y$, the total wavefield $\mathbf{u}$ depends only on $x$ and $z$, which means that the to-be-considered problem is 2D  and can be examined in the sagittal $x-z$ plane.
\begin{figure}[ht]
\begin{center}
\includegraphics[width=0.65\textwidth]{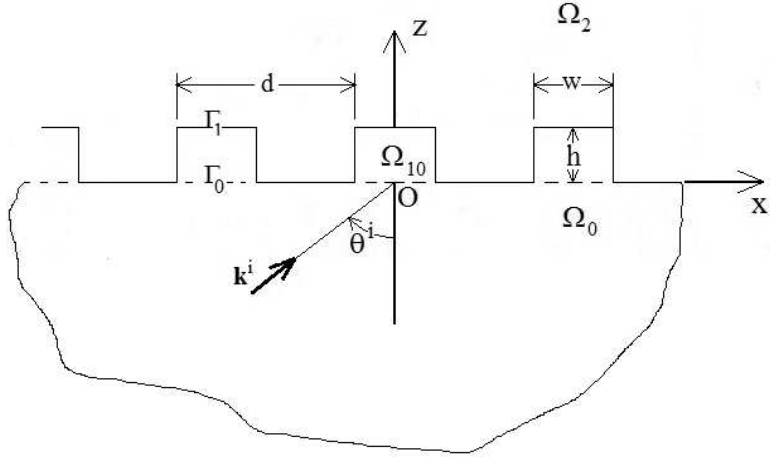}
\caption{Sagittal plane view of the periodically-uneven boundary ('grating' for short) consisting of rectangular protrusions emerging from the ground plane $z=0$/ The medium below the boundary is a homogeneous, isotropic, elastic solid. The central protrusion domain is $\Omega_{10}$ (width $w$, height $h$), its left-hand neighbor is the domain $\Omega_{1-11}$ and its right-hand neighbor is the domain $\Omega_{11}$, etc. The grating, of period $d$, is solicited by a SH plane body wave whose wavevector (lying in the sagittal plane) makes an angle $\theta^{i}$ with the $z$ axis.}
\label{elasticgrating}
\end{center}
\end{figure}
Fig. \ref{elasticgrating} depicts the problem  in the sagittal plane in which: $\Omega_{0}$ is the portion of $\mathcal{L}$ below $z=0$ and $\Omega_{1}=\cup_{n\in\mathbb{Z}}\Omega_{1n}$  the (composite) domain constituted by the remainder of $\mathcal{L}$, with $\Omega_{1n}$ the $n$-th  subdomain (henceforth termed 'protrusion') of rectangular cross section (width $w$ and height $h$). $\Omega_{2}$ ( the sagittal plane trace of $\mathcal{U}$.

The medium in $\mathcal{L}$ is assumed to be non-dispersive over the range of frequencies of interest, and its shear modulus $\mu$ to be real. The shear-wave velocity in this solid is the real quantity $\beta=\sqrt{\frac{\mu}{\rho}}$, with  $\rho$ the mass density.

The  wavevector $\mathbf{k}^{i}$ of the plane wave solicitation lies in the  sagittal plane and is of the form $\mathbf{k}^{i}=(k_{x}^{i},k_{z}^{i})=(k\sin\theta^{i},k\cos\theta^{i})$ wherein  $\theta^{i}$ is the angle of incidence (see fig. \ref{elasticgrating}), and $k=\omega/\beta$. We shall assume $\theta^{i}=0^{\circ}$ in the last part of this study.

The total wavefield is $\mathbf{u}(\mathbf{x},\omega)=\big(0,u(\mathbf{x},\omega),0\big)$ and $u(\mathbf{x},\omega)$ in $\Omega_{l}$ is designated by $u^{[l]}(\mathbf{x},\omega)$. The incident wavefield is
\begin{equation}\label{1-000}
u^{i}(\mathbf{x},\omega)=u^{[0]+}(\mathbf{x},\omega)=a^{[0]+}(\omega)\exp[i(k_{x}^{i}x+k_{z}^{i}z)]~,
\end{equation}
wherein $a^{[0]+}(\omega)$ is the spectral amplitude of the solicitation.

The plane wave nature of the solicitation and the $d$-periodicity of $\mathcal{B}$  entails the quasi-periodicity of the field, whose expression is the Floquet condition
\begin{equation}\label{1-005}
u(x+d,z,\omega)=u(x,z,\omega)\exp(ik_{x}^{i}d)~;~\forall\mathbf{x}\in \Omega_{0}+\Omega_{1}~.
\end{equation}
Consequently, as concerns the response in $\Omega_{1}$, it suffices to examine the field in $\Omega_{10}$.

The boundary-value problem in the space-frequency domain translates to the following relations (in which the superscripts $+$ and $-$ refer to the upgoing and downgoing  waves respectively) satisfied by the total displacement field $u^{[l]}(\mathbf{x};\omega)$ in $\Omega_{l}$:
\begin{equation}\label{1-010}
u^{[l]}(\mathbf{x},\omega)=u^{[l]+}(\mathbf{x},\omega)+u^{[l]-}(\mathbf{x},\omega)~;~l=0,1~,
\end{equation}
\begin{equation}\label{1-020}
u_{,xx}^{[l]}(\mathbf{x},\omega)+u_{,zz}^{[l]}(\mathbf{x},\omega)+k^{2}u^{[l]}(\mathbf{x},\omega)=0~;~\mathbf{x}\in \Omega_{l}~;~l=0,1~.
\end{equation}
\begin{equation}\label{1-030}
\mu u_{,z}^{[1]}(x,h,\omega)=0~;~\forall x\in [-w/2,w/2]~,
\end{equation}
\begin{equation}\label{1-033}
\mu u_{,z}^{[0]}(x,0,\omega)=0~;~\forall x\in [-d/2,w/2]\cup [w/2,d/2]~,
\end{equation}
\begin{equation}\label{1-035}
\mu u_{,x}^{[1]}(\pm w/2,z,\omega)=0~;~\forall z\in [0,h]~,
\end{equation}
\begin{equation}\label{1-040}
u^{[0]}(x,0,\omega)-u^{[1]}(x,0,\omega)=0~;~\forall x \in [-w/2,w/2]~,
\end{equation}
\begin{equation}\label{1-050}
\mu u_{,z}^{[0]}(x,0,\omega)-\mu u_{,z}^{[1]}(x,0,\omega)=0~;~\forall x \in [-w/2,w/2]~,
\end{equation}
wherein   $u_{,\zeta}$ ($u_{,\zeta\zeta}$) denotes the first (second) partial derivative of $u$ with respect to $\zeta$. Eq. (\ref{1-020}) is the space-frequency  SH wave equation, (\ref{1-030})-(\ref{1-035}) the stress-free boundary conditions, (\ref{1-040}) the expression of continuity of displacement across the junction  between the $\Omega_{0}$ and the central block,  and (\ref{1-050}) the expression of continuity of stress across this junction.

Since $\Omega_{0}$ is of half-infinite extent, the field therein must obey the radiation condition
\begin{equation}\label{1-060}
u^{[0]-}(\mathbf{x},\omega)\sim \text{outgoing waves}~;~\mathbf{x}\rightarrow\infty~.
\end{equation}
%
\subsection{Field representations via domain decomposition and separation of variables (DD-SOV)}\label{sov}
As the preceding descriptions emphasize, it is natural to decompose the domain below the stress-free surface into the central protrusive domain above the ground and the half space domain beneath the ground.

Applying the SOV technique, The Floquet condition, and the radiation condition gives rise, in the  lower domain, to the field representation:
\begin{equation}\label{2-010}
u^{[0]\pm}(\mathbf{x},\omega)=\sum_{n\in\mathbb{Z}}a_{n}^{[0]\pm}(\omega)\exp[i(k_{xn}^{[0]}x\pm k_{zn}^{[0]}z)]~,
\end{equation}
wherein:
\begin{equation}\label{2-012}
k_{xn}^{[0]}=k_{x}^{i}+\frac{2n\pi}{d}~,
\end{equation}
\begin{equation}\label{2-020}
k_{zn}^{[0]}=\sqrt{k^{2}-\left(k_{xn}^{[0]}\right)^{2}}~~;~~\Re k_{zn}^{[0]}\ge 0~~,~~\Im k_{zn}^{[0]}\ge 0~~\omega>0~,
\end{equation}
and, on account of (\ref{1-000}),
\begin{equation}\label{2-030}
a_{n}^{[0]+}(\omega)=a^{[0]+}(\omega)~\delta_{n0}~,
\end{equation}
with $\delta_{n0}$ the Kronecker delta symbol.

In the central protrusion, the SOV, together with the free-surface boundary conditions (\ref{1-030}), (\ref{1-035}), lead to
\begin{equation}\label{2-040}
u^{[1]\pm}(\mathbf{x},\omega)=\frac{1}{2}\sum_{m=0}^{\infty}a_{m}^{[1]}(\omega)\cos[k_{xm}^{[1]}(x+w/2)]\exp[\pm k_{zm}^{[1]}(z-h)]~,
\end{equation}
in which
\begin{equation}\label{2-050}
k_{xm}^{[1]}=\frac{m\pi}{w}~,
\end{equation}
\begin{equation}\label{2-060}
k_{zm}^{[1}=\sqrt{k^{2}-\big(k_{xm}^{[1]}\big)^{2}}~~;~~\Re k_{zm}^{[1]}\ge 0~~,~~\Im k_{zm}^{[1]}\ge 0~~\omega>0~.
\end{equation}
%
\subsection{Exact solutions for the unknown coefficients}
Eqs. (\ref{1-033}) and (\ref{1-050})  entail
\begin{equation}\label{3-003}
\mu\int_{-d/2}^{d/2}u^{[0]}_{,z}(x,0,\omega)\exp(-ik_{xj}^{[0]}x)\frac{dx}{d}=
\mu\int_{-w/2}^{w/2}u^{[01}_{,z}(x,0,\omega)\exp(-ik_{xj}^{[0]}x)\frac{dx}{d}~;~\forall j=0,\pm 1,\pm 2,....~,
\end{equation}
which, on account of the SOV field representations and the identity
\begin{equation}\label{3-005}
\int_{-d/2}^{d/2}\exp\left[i\left(k_{xn}^{[0]}-k_{xj}^{[0]}\right)x\right]\frac{dx}{d}=\delta_{nj}~,
\end{equation}
($\delta_{nj}$ is the Kronecker delta) yields
\begin{equation}\label{3-010}
a_{j}^{[0]-}=a_{j}^{[0]+}-\frac{w}{2id}\frac{1}{k_{zj}^{[0]}}
\sum_{m=0}^{\infty}a_{m}^{[1]}k_{zm}^{[1]}\sin\left(k_{zm}^{[1]}h\right)E_{jm}^{-}~;~\forall j=0,\pm 1,\pm 2,....
~,
\end{equation}
wherein
\begin{multline}\label{3-040}
E_{jm}^{\pm}=\int_{-w/2}^{w/2}\exp\left(\pm ik_{xj}^{[0]}x\right)\cos\left[k_{xm}^{[1]}(x+w/2)\right]\frac{dx}{w/2}=\\
i^{m}\left\{\text{sinc}\left[\left(\pm k_{xj}^{[0]}+k_{xm}^{[1]}\right)w/2\right]+(-1)^{m}\text{sinc}\left[\left(\pm k_{xj}^{[0]}-k_{xm}^{[1]}\right)w/2\right]\right\}
~,
\end{multline}
with sinc$(\zeta)=\frac{\sin\zeta}{\zeta}$ and sinc(0)=1.

Eq. (\ref{1-040})  entails
\begin{multline}\label{3-050}
\int_{-w/2}^{w/2}u^{[[0]}(x,0,\omega)\cos\left[k_{xl}({[1]}(x+w/2)\right]\frac{dx}{w/2}=\\
\int_{-w/2}^{w/2}u^{[[1]}(x,0,\omega)\cos\left[k_{xl}({[1]}(x+w/2)\right]\frac{dx}{w/2}~;~\forall l=0,1,2,....~,
\end{multline}
which, on account of the SOV field representations, and the identity
\begin{equation}\label{3-055}
\int_{-w/2}^{w/2}\cos\left[k_{xm}^{[1]}(x+w/2)\right]\cos\left[k_{xl}^{[1]}(x+w/2)\right]=\frac{2}{\epsilon_{l}}\delta_{lm}~,
\end{equation}
with $\epsilon_{l}$ the Neumann symbol (=1 for $l=0$ and =2 for $l>0$), enables us to find
\begin{equation}\label{3-060}
a_{l}^{[1]}=\left(\frac{\epsilon_{l}}{2\cos\left(k_{zl}^{[1]}h\right)}\right)
\sum_{n-\infty}^{\infty}\left(a_{n}^{[0]+}+a_{n}^{[0]-}\right)E_{nl}^{+}~;~\forall l=0,1,2,....
\end{equation}
We thus have at our disposal two coupled expressions (i.e., (\ref{3-010}) and (\ref{3-060}) which make it possible to determine the two sets of unknowns $\{a_{n}^{[0]-}\}$, $\{a_{n}^{[1]}\}$. Note that the number of members of each of these sets is infinite which is the fundamental source of  complexity of the problem at hand and the principal reason why one should strive to simplify the theoretical analysis. This will be done in a later section.
\subsection{Linear system for the  set of unknown coefficients}
Iinserting (\ref{3-010}) into(\ref{3-060})  yields, after the summation interchange, the system of linear equations:
\begin{equation}\label{4-040}
\sum_{m=0}^{\infty}X_{lm}Y_{m}=Z_{l}~;~\forall l=0,1,2,....~,
\end{equation}
wherein
\begin{equation}\label{4-045}
Y_{m}=a_{m}^{[1]}~,~~Z_{l}=a^{[0]+}\epsilon_{l}E_{0l}^{+}~,
\end{equation}
\begin{equation}\label{4-050}
X_{lm}=\delta_{lm}\cos\left(k_{zm}^{[1]}h\right)+\frac{w}{2id}\frac{\epsilon_{l}}{2}k_{zm}^{[1]}\sin\left(k_{zm}^{[1]}h\right)\Sigma_{lm}~~,~~
\Sigma_{lm}=\sum_{n=-\infty}^{\infty}\frac{1}{k_{zn}^{[0]}}E_{nl}^{+}E_{nm}^{-}~.
\end{equation}
Once the $Y_{m}=a_{m}^{[1]}$ are determined they can be inserted into (\ref{3-010}) to determine the $a_{j}^{[0]-}$, i.e.,
\begin{equation}\label{4-055}
a_{j}^{[0]-}=a_{j}^{[0]+}-\frac{w}{2id}\frac{1}{k_{zj}^{[0]}}
\sum_{m=0}^{\infty}Y_{m}k_{zm}^{[1]}\sin\left(k_{zm}^{[1]}h\right)E_{jm}^{-}~;~\forall j=0,\pm 1,\pm 2,....
~,
\end{equation}

Until now everything has been rigorous provided the equations in the statement of the boundary-value problem are accepted as the true expression of what is involved in the  elastic wave response of our grating and certain summation interchanges are valid. In order to actually solve for the sets $\{a_{n}^{[0]-}\}$ and $\{a_{m}^{[1]}\}$ (each of whose populations is  considered to be infinite at this stage) we must resort either to numerics or to approximations.
\subsection{Numerical issues concerning the  system of equations for $\{a_{m}^{[1]}\}$}
We strive to obtain numerically the set $\{a_{m}^{[1]}\}$ from the linear system of equations (\ref{4-040}). Once this set is found, it is introduced into (\ref{3-010}) to obtain the set $\{a_{n}^{[0]-}\}$.  When all these coefficients (we mean those whose values depart significantly from zero) are found, they enable the computation of the elastic wave response (i.e., the displacement field) in all the subdomains of the configuration via (\ref{1-000}), (\ref{1-010}), (\ref{2-010}), (\ref{2-040}).

Concerning the resolution of the infinite system of linear equations (\ref{4-040}), the procedure is basically to replace it by the finite system of linear equations
\begin{equation}\label{5-010}
\sum_{m=0}^{M}X^{(M)}_{lm}Y_{m}^{(M)}=Z_{l}~;~l=0,1,2,...M~,
\end{equation}
in which $X^{(M)}_{lm}$ signifies that the series in $X_{lm}$ is limited to the terms $n=0,\pm 1,...,\pm M$,  and to increase $M$ so as to generate the sequence of numerical solutions $\{Y_{m}^{(0)}\}$, $\{Y_{m}^{(1)},Y_{m}^{(2)}\}$,....until the values of the first few members of  of these sets stabilize and the remaining members become very small (this is the so-called 'reduction method' \cite{ri13} of resolution of an infinite system of linear equations).

Note that to each $Y_{m}^{(M)}=a_{m}^{[1](M)}$ is associated $a_{m}^{[0]-(M)}$ via (\ref{4-055}), i.e.,
\begin{equation}\label{5-015}
a_{j}^{[0]-(M)}=a_{j}^{[0]+}-\frac{w}{2id}\frac{1}{k_{zj}^{[0]}}
\sum_{m=0}^{M}Y_{m}^{(M)}k_{zm}^{[1]}\sin\left(k_{zm}^{[1]}h\right)E_{jm}^{-}~;~\forall j=0,\pm 1,\pm 2,....\pm M
~,
\end{equation}

The so-obtained numerical solutions (it being implicit that $Y_{m}^{(M)}=a^{[1](M)}=0~;~m>M$ and $a_{j}^{[0]-(M)}~;~|j|>M$), which for all practical purposes can be considered as 'exact' for sufficiently-large $M$ (of the order of 25 for the range of frequencies and uneveness parameters considered herein) and which are in agreement with numerical results obtained by a finite element method \cite{gr05,gw08}, constitute the reference by which we shall measure the accuracy of the approximate solutions of the next section.

From this point on, we assume that the  plane wave is normally-incident (i.e., $\theta^{i}=0^{\circ}$) onto the uneven boundary $\mathcal{B}$. The consequences of this are
\begin{equation}\label{5-025}
k_{xj}^{[0]}=2j\pi/d~~,~~k_{x0}^{[0]}=0~~,~~k_{z0}^{[0]}=k_{z0}^{[1]}=k.
~,
\end{equation}
%
\subsubsection{Dependence of reduction method solutions on $M$ for varying frequency and various $w$}\label{numMredfw}
\begin{figure}[ht]
\begin{center}
\includegraphics[width=0.75\textwidth]{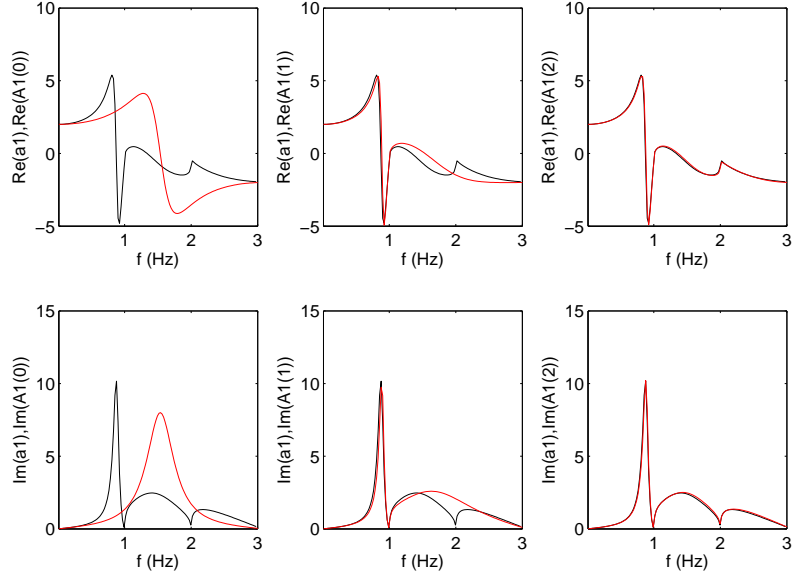}
\caption{The black curves represent the reference spectra $a_{0}^{[1]}(f)$ and the red curves the approximate spectra $a_{0}^{[1](M)}(f)$. The upper (lower) panels correspond to the real (imaginary) parts of these functions. The left-hand, middle and right-hand panels are for $M=0,1,2$ respectively. Case $w=430~m$,  $h=280~m$, }
\label{redfw-01}
\end{center}
\end{figure}

\begin{figure}[ptb]
\begin{center}
\includegraphics[width=0.75\textwidth]{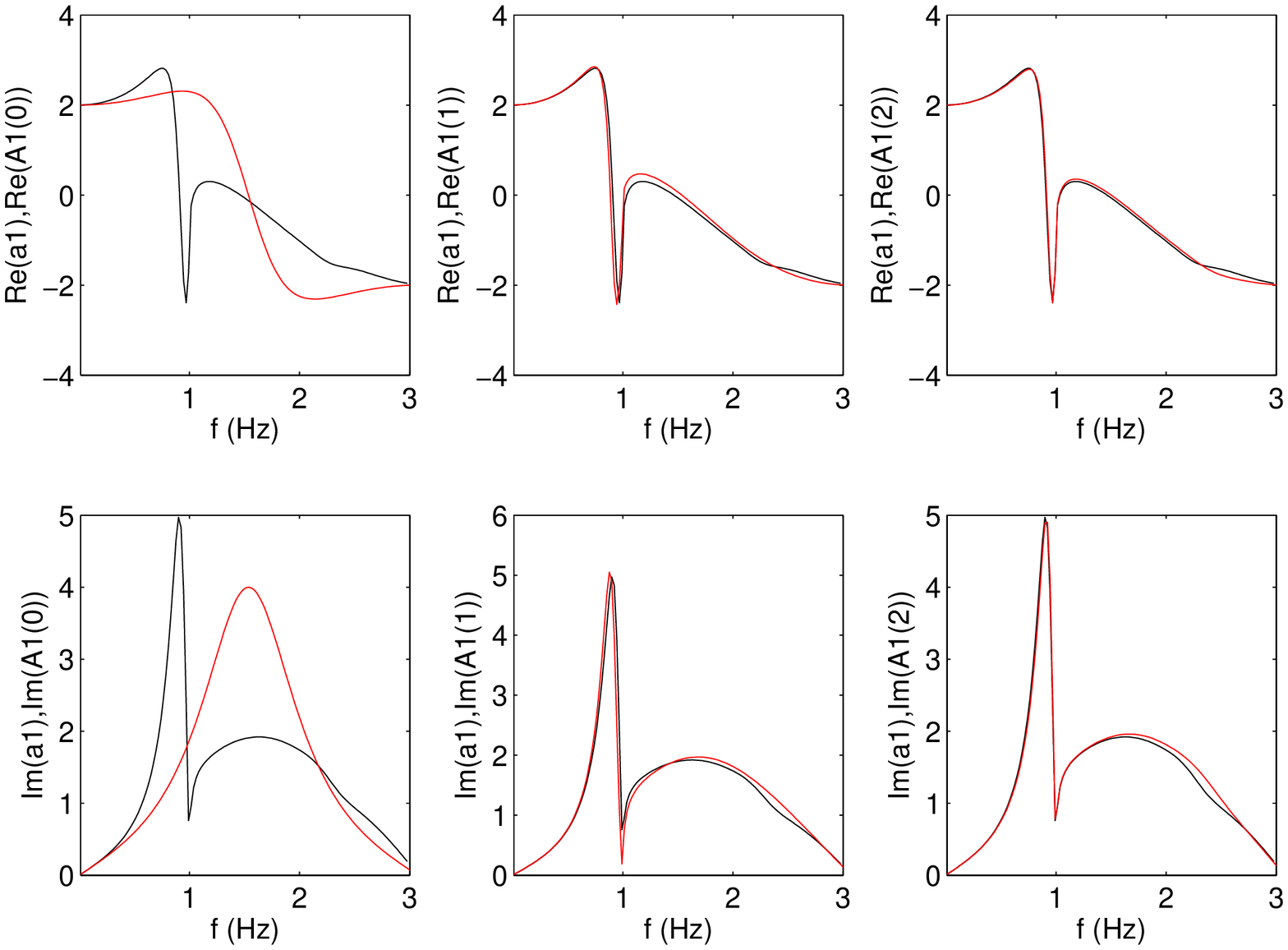}
\caption{Same as fig. \ref{redfw-01} except that $w=860~m$,  $h=280~m$.}
\label{redfw-02}
\end{center}
\end{figure}

\begin{figure}[ptb]
\begin{center}
\includegraphics[width=0.75\textwidth]{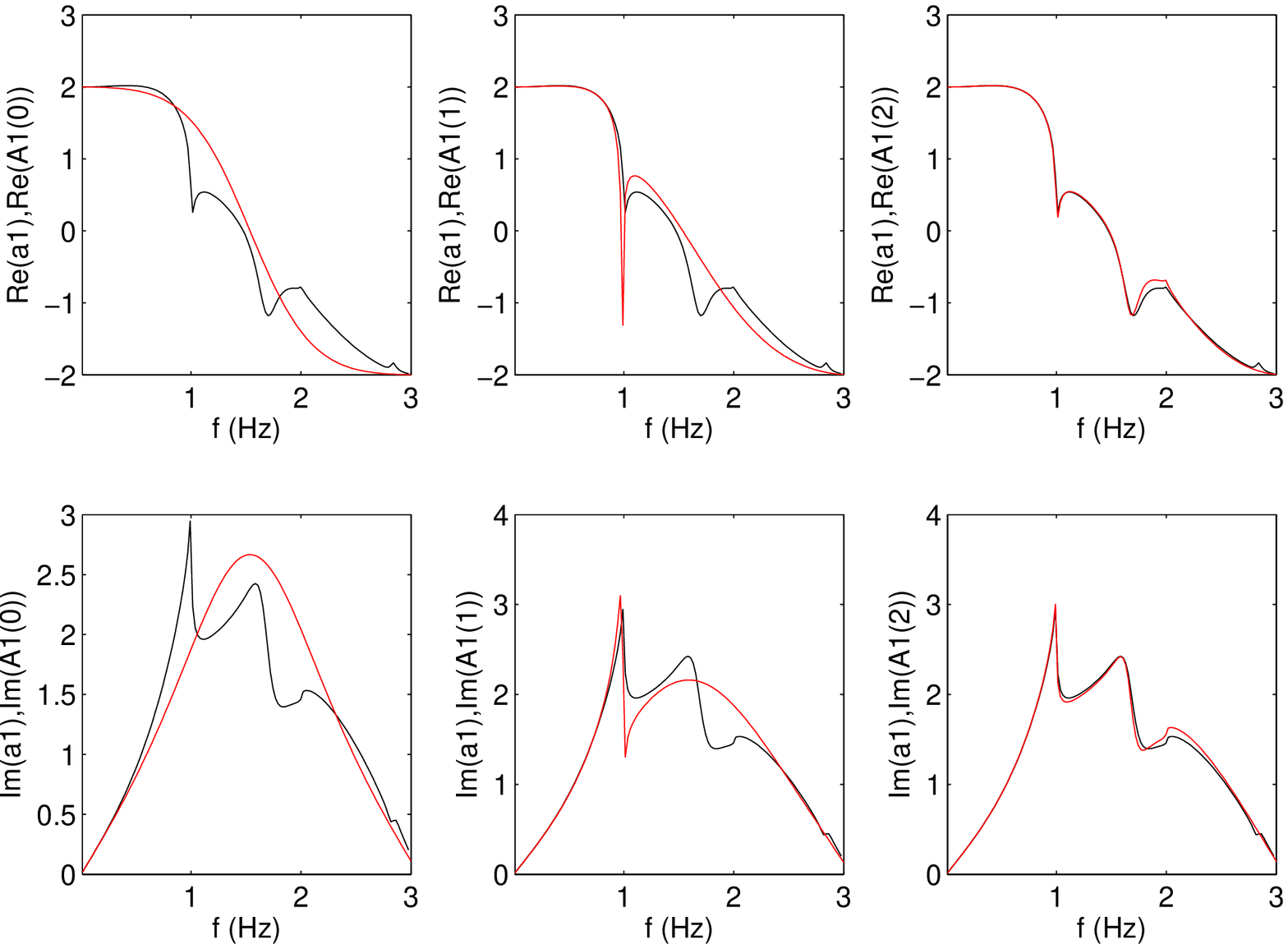}
\caption{Same as fig. \ref{redfw-01} except that $w=1290~m$,  $h=280~m$.}
\label{redfw-03}
\end{center}
\end{figure}

Figs. \ref{redfw-01}-\ref{redfw-03} tell us how the reduction method solutions evolve with $M$ for varying $f$ and various $w$. In all these figures, $\theta^{i}=0^{\circ}$, $d=1720~m$, $\beta^{[0]}=17200~ms^{-1}$, $\mu^{[0]}=1\times 10^{9}~Pa$,  $\beta^{[1]}=1720~ms^{-1}$, $\mu^{[1]}=1\times 10^{9}~Pa$ and the reference solutions (black) curves are obtained for $M=25$.
\clearpage
\newpage
A notable feature of these figures is that the larger is $M$, the better is the agreement with the reference results (obtained for large $M$) at a given (especially close to resonant and/or high) frequency, or stated otherwise: the closer to a resonant $f$ or the higher the frequency, the greater $M$ must be for the $M$-th order solution to agree with the reference solution. Note that the required value of $M$ is not a linear function of frequency $f$.
\subsubsection{Dependence of reduction method solutions on $M$ for varying frequency and various $h$}\label{numMredfh}
\begin{figure}[ht]
\begin{center}
\includegraphics[width=0.75\textwidth]{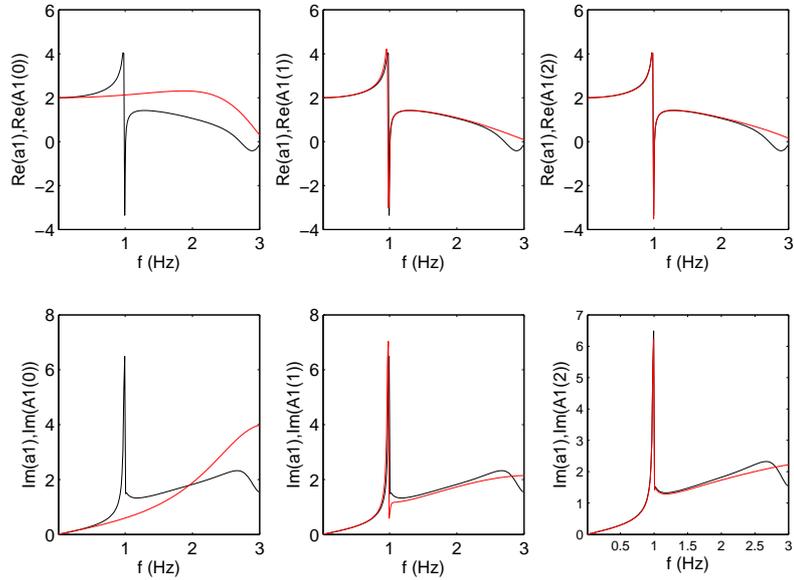}
\caption{Same as fig. \ref{redfw-01} except that $w=860~m$,  $h=140~m$, }
\label{redfh-01}
\end{center}
\end{figure}

\begin{figure}[ptb]
\begin{center}
\includegraphics[width=0.75\textwidth]{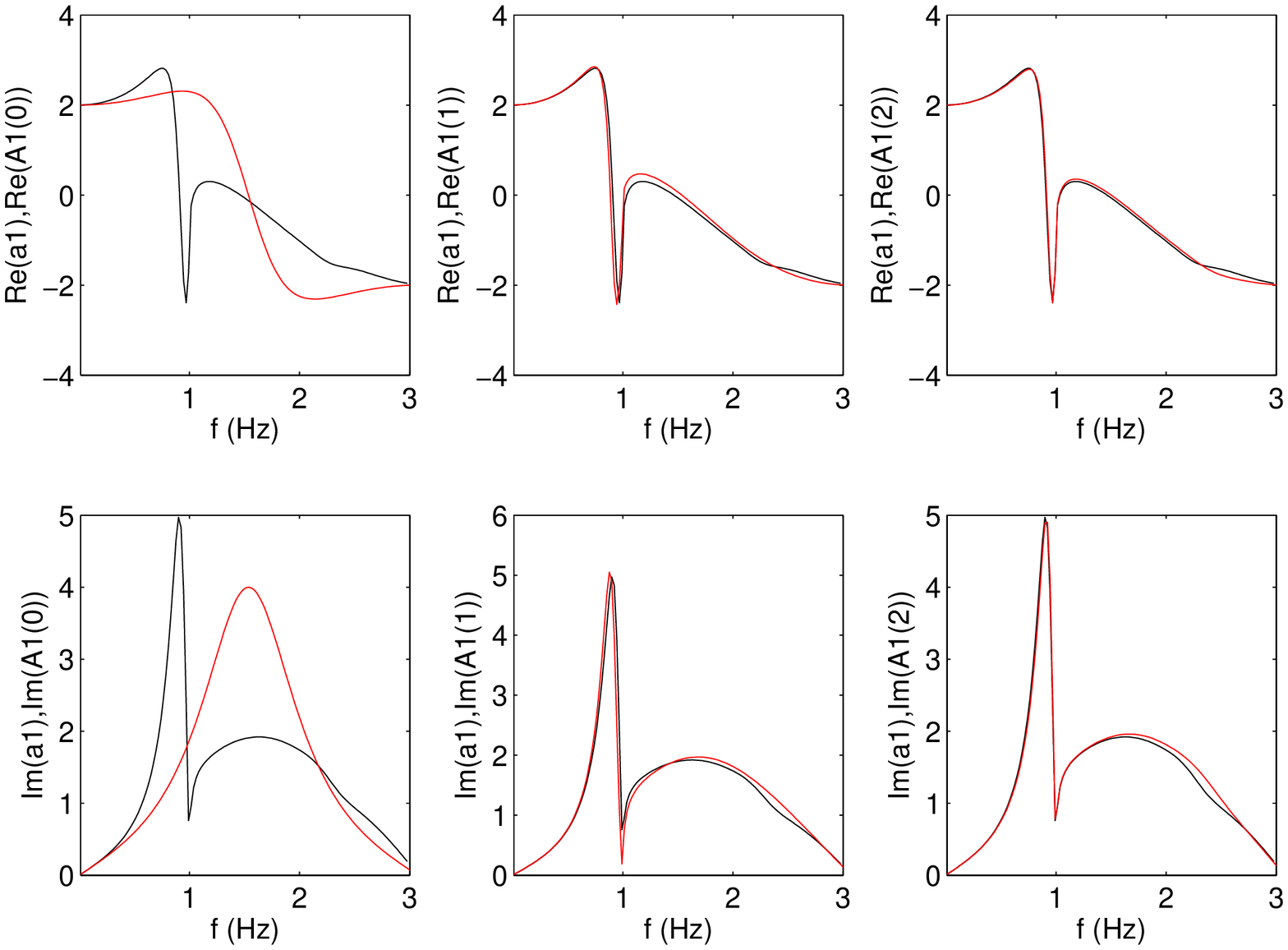}
\caption{Same as fig. \ref{redfw-01} except that $w=860~m$,  $h=280~m$.}
\label{redfh-02}
\end{center}
\end{figure}

\begin{figure}[ptb]
\begin{center}
\includegraphics[width=0.75\textwidth]{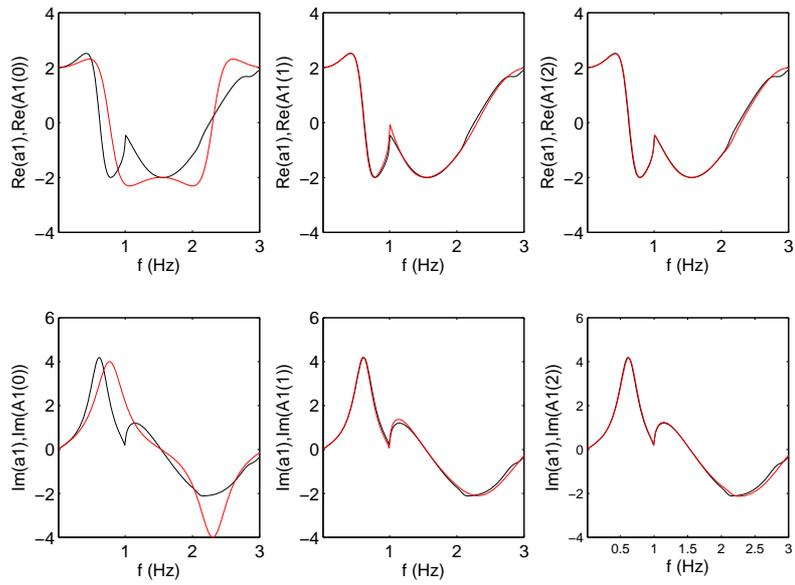}
\caption{Same as fig. \ref{redfw-01} except that $w=860~m$,  $h=560~m$.}
\label{redfh-03}
\end{center}
\end{figure}

Figs. \ref{redfh-01}-\ref{redfh-03} tell us how the reduction method solutions evolve with $M$ for varying $f$ and various $w$. In all these figures, $\theta^{i}=0^{\circ}$, $d=1720~m$, $\beta^{[0]}=17200~ms^{-1}$, $\mu^{[0]}=1\times 10^{9}~Pa$,  $\beta^{[1]}=1720~ms^{-1}$, $\mu^{[1]}=1\times 10^{9}~Pa$ and the reference solutions (black) curves are obtained for $M=25$.
\clearpage
\newpage
A notable feature of these figures is that the larger is $M$, the better is the agreement with the reference results (obtained for large $M$) at a given (especially close to resonant and/or high) frequency, or stated otherwise: the closer to a resonant $f$ or the higher the frequency, the greater $M$ must be for the $M$-th order solution to agree with the reference solution. Note that the required value of $M$ is not a linear function of frequency $f$.

\section{The $M=0$ approximation of the response of the uneven boundary}\label{Meq0}\label{Meq0}
When $M=0$, the consequence of  (\ref{4-040}) is
\begin{equation}\label{6-020}
Y_{0}^{(0)}=\frac{Z_{0}}{X_{00}^{(0)}}~,
\end{equation}
whence
\begin{equation}\label{6-025}
Y_{0}^{(0)}=a_{0}^{[1]}=a^{[0]+}\left[\frac{E_{00}^{+}}{\cos\left(k_{z0}^{[1]}h\right)+
\frac{w}{4id}k_{z0}^{[1]}\sin\left(k_{z0}^{[1]}h\right)\Sigma_{00}}\right]
~,
\end{equation}
or (taking account of (\ref{5-025}))
\begin{equation}\label{6-030}
Y_{0}^{(0)}=a_{0}^{[1](0)}=a^{[0]+}\left[
\frac{2}
{
\cos\left(kh\right)+
\frac{w}{id}
\sin\left(kh\right)
}
\right]
~,
\end{equation}
whence
\begin{equation}\label{6-040}
a_{0}^{[0]-(0)}=a^{[0]+}\left[
\frac{
\cos\left(kh\right)-
\frac{w}{id}
\sin\left(kh\right)}
{\cos\left(kh\right)+
\frac{w}{id}
\sin\left(kh\right)
}
\right]
~.
\end{equation}

A word is here in order about the possibility of resonance showing up in these response functions. Resonances, typically those associated with the excitation of Love modes \cite{ejp57}, occur (i.e., at a discrete set of frequencies) when the denominator in the response functions are equal or very nearly equal  to zero, therefore leading to infinite or very large response. For this to occur, while assuming, as we have done in this study that the medium in $\mathcal{L}$ is lossless, would require that the  $\sin$ term in the denominators of (\ref{6-030}) and (\ref{6-040}) be real and negative in relation to the $\cos$ term, but this is impossible because the factor $w/id$ multiplying $\sin$ is imaginary. It follows that the $M=0$ approximation of the uneven boundary response cannot account for resonant behavior, which fact was already observed in the numerical results presented in sects. \ref{numMredfw}-\ref{numMredfh}. We shall return to this issue in sect. \ref{comp}.
\section{Relation of the approximate $M=0$ solution to the exact solution of another problem}
The $M=0$  approximate solution (\ref{6-030})-(\ref{6-040}) of the uneven boundary problem
resembles the exact solution of a problem in which the uneven boundary is replaced by a flat-faced layer occupied by a linear, homogeneous, isotropic solid, the  solicitation being the same  (as regards the polarization of the plane body wave) as for the uneven boundary. To substantiate this assertion, we first derive the solution of the layer problem.
\subsection{Description of the problem of the response to a plane wave of a homogeneous layer above a half space}
The bottom flat face of the layer (in firm contact with the underlying solid medium) occupies the entire plane $z=0$ and the upper flat face of the layer occupies the entire $z=H$ plane. The half-space above the layer is occupied by the vacuum and the  half-space below the layer by a linear, homogeneous, isotropic, lossless, non-dispersive solid.

The elastic wave solicitation is the same (i.e, as regards the polarization of the plane body wave) as in the uneven boundary  problem. The  wavefield associated with the elastic wave solicitation is $\mathbf{U}^{i}=(U_{x}^{i},U_{y}^{i},U_{z}^{i})=(0,U^{i}(\mathbf{x},\omega),0)$.

Since neither the incident wavefield nor the geometric and compositional features of the configuration  depend on $y$, the total wavefield $\mathbf{U}=(0,U,0)$ depends only on $x$ and $z$, which means that the to-be-considered problem is 2D  and can be examined in the sagittal $x-z$ plane.
\begin{figure}[ht]
\begin{center}
\includegraphics[width=0.65\textwidth]{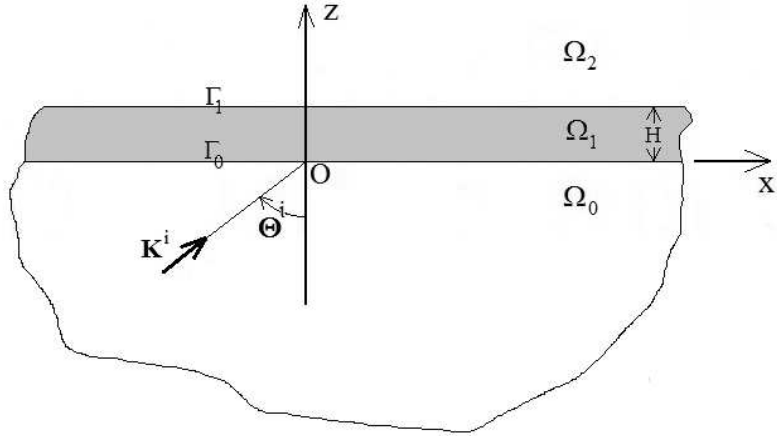}
\caption{Sagittal plane view of the   configuration  comprising a  homogeneous layer in firm contact with the underlying homogeneous solid across $z=0$. The half-space below $z=0$ is $\Omega_{0}$,  the layer domain is $\Omega_{1}$ (height $H$) and the half-space above the layer is $\Omega_{2}$. The configuration is solicited by a $SH$ plane body wave whose wavevector (lying in the sagittal plane) makes an angle $\Theta^{i}$ with the $z$ axis.}
\label{layerlikegrating}
\end{center}
\end{figure}
Fig. \ref{layerlikegrating} depicts the problem  in the sagittal plane in which: $\Omega_{0}$ is the half-space domain occupied by a  linear, homogeneous, isotropic, lossless, non-dispersive solid,  $\Omega_{1}$  the domain of the layer occupied by another  linear, homogeneous, isotropic material which might be lossy and/or dispersive,  and $\Omega_{2}$ the  half-space above the layer occupied by the vacuum.

The shear moduli of the lower medium and layer are $M^{[0]}$ and $M^{[1]}$ respectively, with  $M^{[0]}$  positive real and $M^{[1]}$ generally-complex.  The shear-wave velocities in the lower medium and layer are $B^{[0]}$ and $B^{[1]}$ respectively, with  $B^{[0]}$  positive real and $B^{[1]}$ generally-complex.

The  wavevector $\mathbf{K}^{i}$ of the plane wave solicitation lies in the  sagittal plane and is of the form $\mathbf{K}^{i}=(K_{x}^{i},K_{z}^{i})=(K^{[0]}\sin\Theta^{i},K^{[0]}\cos\Theta^{i})$ wherein  $\Theta^{i}$ is the  angle of incidence (see fig. \ref{layerlikegrating}), and $K^{[l]}=\omega/B^{[l]}$.

The total wavefield $U(\mathbf{x},\omega)$ in $\Omega_{l}$ is designated by $U^{[l]}(\mathbf{x},\omega)$. The incident wavefield is
\begin{equation}\label{7-020}
U^{i}(\mathbf{x},\omega)=U^{[0]+}(\mathbf{x},\omega)=A^{[0]+}(\omega)\exp[i(K_{x}^{i}x+K_{z}^{i}z)]~,
\end{equation}
wherein $A^{[0]+}(\omega)$ is the spectral amplitude of the solicitation.
\subsection{The boundary-value problem of the response of the layer/halfspace to a plane wave}
The boundary-value problem in the space-frequency domain translates to the following relations  satisfied by the total displacement field $U^{[l]}(\mathbf{x};\omega)$ in $\Omega_{l}$:
\begin{equation}\label{7-030}
U^{[l]}(\mathbf{x},\omega)=U^{[l]+}(\mathbf{x},\omega)+U^{[l]-}(\mathbf{x},\omega)~;~l=0,1~,
\end{equation}
\begin{equation}\label{7-040}
U_{,xx}^{[l]}(\mathbf{x},\omega)+U_{,zz}^{[l]}(\mathbf{x},\omega)+(K^{[l]})^{2}U^{[l]}(\mathbf{x},\omega)=0~;~\mathbf{x}\in \Omega_{l}~;~l=0,1~.
\end{equation}
\begin{equation}\label{7-050}
M^{[1]}U_{,z}^{[1]}(x,H,\omega)=0~;~\forall x\in \mathbb{R}~,
\end{equation}
\begin{equation}\label{7-060}
U^{[0]}(x,0,\omega)-U^{[1]}(x,0,\omega)=0~;~\forall x \in \mathbb{R},
\end{equation}
\begin{equation}\label{7-070}
M^{[0]}U_{,z}^{[0]}(x,0,\omega)-M^{[1]}U_{,z}^{[1]}(x,0,\omega)=0~;~\forall x \in \mathbb{R}~.
\end{equation}
\begin{equation}\label{7-080}
U^{[0]-}(\mathbf{x},\omega)\sim \text{outgoing waves}~;~\mathbf{x}\rightarrow\infty~.
\end{equation}
%
\subsection{DD-SOV field representations}
Applying the DD-SOV technique, and the radiation condition gives rise, in the  lower domain, to the field representation:
\begin{equation}\label{7-090}
U^{[0]\pm}(\mathbf{x},\omega)=A^{[0]\pm}(\omega)\exp[i(K_{x}^{[0]}x\pm K_{z}^{[0]}z)]~,
\end{equation}
wherein:
\begin{equation}\label{7-100}
K_{x}^{[0]}=K_{x}^{i}~,
\end{equation}
\begin{equation}\label{7-120}
K_{z}^{[0]}=\sqrt{\big(K^{[0]}\big)^{2}-\left(K_{x}^{[0]}\right)^{2}}~~;~~\Re K_{z}^{[0]}\ge 0~~,~~\Im K_{z}^{[0]}\ge 0~~\omega>0~.
\end{equation}

In the layer, the SOV, together with the free-surface boundary condition (\ref{7-050}),  lead to
\begin{equation}\label{7-140}
U^{[1]}(\mathbf{x},\omega)=A^{[1]}(\omega)\exp\left[iK_{x}^{[1]}x\right]\cos\left[K_{z}^{[1]}(z-H)\right]~,
\end{equation}
in which
\begin{equation}\label{7-150}
K_{x}^{[1]}=K_{x}^{[0]}=K_{x}^{i}~,
\end{equation}
\begin{equation}\label{7-160}
K_{z}^{[1]}=\sqrt{\big(K^{[1]}\big)^{2}-\big(K_{x}^{[1]}\big)^{2}}~~;~~\Re K_{z}^{[1]}\ge 0~~,~~\Im K_{z}^{[1]}\ge 0~~\omega>0~.
\end{equation}
%
\subsection{Exact solution for the unknown coefficients}
The introduction of the field representations into  (\ref{7-060})-(\ref{7-070}) yields the two equations
\begin{equation}\label{7-170}
A^{[0]+}+A^{[0]-}=A^{[1]}\cos\left(-K_{z}^{[1]}H\right)~,
\end{equation}
\begin{equation}\label{7-180}
iM^{[0]}K_{z}^{[0]}\left(A^{[0]+}-A^{[0]-}\right)=-M^{[1]}K_{z}^{[1]}A^{[1]}\sin\left(-K_{z}^{[1]}H\right)~,
\end{equation}
the exact solution of which is:
\begin{equation}\label{7-190}
A^{[1]}=A^{[0]+}\left[\frac{2}
{\cos\left(K_{z}^{[1]}H\right)+\frac{M^{[1]}K_{z}^{[1]}}{iM^{[0]}K_{z}^{[0]}}\sin\left(K_{z}^{[1]}H\right)}\right]~.
\end{equation}
\begin{equation}\label{7-195}
A^{[0]-}=A^{[0]+}\left[\frac{\cos\left(K_{z}^{[1]}H\right)-\frac{M^{[1]}K_{z}^{[1]}}{iM^{[0]}K_{z}^{[0]}}\sin\left(K_{z}^{[1]}H\right)}
{\cos\left(K_{z}^{[1]}H\right)+\frac{M^{[1]}K_{z}^{[1]}}{iM^{[0]}K_{z}^{[0]}}\sin\left(K_{z}^{[1]}H\right)}\right]~,
\end{equation}
%
\subsection{Comparison of the approximate uneven boundary problem solution to the exact layer problem solution}\label{comp}
The comparison of (\ref{6-030})-(\ref{6-040}) with (\ref{7-190})-(\ref{7-195}) shows that the zeroth-order approximate solution of the uneven boundary problem is structurally-similar to the exact solution of the homogeneous layer problem. This suggests that the layer response is 'equivalent' to the approximate uneven boundary response  when
\begin{equation}\label{7-200}
A^{[1]}=a_{0}^{[1](0)}~,
\end{equation}
which, of course, implies
\begin{equation}\label{7-205}
A^{[0]-}=a_{0}^{[0]-(0)}~.
\end{equation}
The translation of this  equivalence is a series of relations between the parameters of the layer and their counterparts in the grating. The most obvious of these relations are three in number.

Henceforth, we shall assume $\Theta^{i}=0$ which means that the solicitation in both problems is that of a normally-incident plane body wave.
\subsubsection{First series of relations between the layer parameters and the grating parameters}\label{first}
Eq. (\ref{7-200}) is satisfied  provided:
\begin{equation}\label{7-210}
\begin{array}{l}
1a)~B^{[0]}=\beta~\Rightarrow~K^{[0]}=k~,~K_{x}^{[0]}=k_{x0}^{[0]}=0~,~K_{z}^{[0]}=k_{z0}^{[0]}=k~,\\
1b)~B^{[1]}=\beta~\Rightarrow~K^{[1]}=k~,~K_{z}^{[1]}=k_{z0}^{[1]}=k~,\\
1c)~A^{[0]+}=a^{[0]+}~,\\
1d)~H=h~,\\
1e)~M^{[0]}=\mu~,\\
1f)~M^{[1]}=\mu\frac{w}{d}~.
\end{array}
\end{equation}
Note that {\it 1a)-1f)} constitute explicit solutions for all six layer problem parameters. Also, note that all the parameters of the layer problem, just as those of the uneven boundary problem, do not depend  on the frequency.
 It turns out (see figs. \ref{resMe0-01}-\ref{resMe0-03} hereafter and \cite{wi18a,wi18c}) that these effective media account very well for the response of the uneven boundary at very low frequencies, and in any case, before the onset of resonances,
\subsubsection{Second series of relations between the layer parameters and the grating parameters}\label{second}
Eq. (\ref{7-200}) is satisfied  provided:
\begin{equation}\label{7-220}
\begin{array}{l}
2a)~B^{[0]}=\beta~\Rightarrow~K^{[0]}=k~,~K_{x}^{[0]}=k_{x0}^{[0]}=0~,~K_{z}^{[0]}=k_{z0}^{[0]}=k~,\\
2b)~H=h~,\\
2c)~A^{[0]+}=a^{[0]+}~,\\
2d)~M^{[0]}=\mu~,\\
2e)~M^{[1]}=\mu\frac{w}{d}~\\
2f)~K_{z}^{[1]} \text{ solution ($\ne~k$) of } \cos\left(K_{z}^{[1]}H\right)+\frac{M^{[1]}K_{z}^{[1]}}{iM^{[0]}K_{z}^{[0]}}\sin\left(K_{z}^{[1]}H\right)=\\
~~~\cos\left(kh\right)+\frac{w}{id}\sin\left(kh\right)~\text{and~}B^{[1]}=\omega/ K^{[1]}=\omega/K_{z}^{[1]}~.\\
\end{array}
\end{equation}
Note that now only {\it 2a)-2e)} constitute explicit solutions for five of the layer problem parameters whereas the obtention of $B^{[1]}$ requires solving a nonlinear equation for each frequency and the solution of this equation is not unique.
\subsubsection{Third series of relations between the layer parameters and the grating parameters}\label{third}
Eq. (\ref{7-200}) is satisfied  provided:
\begin{equation}\label{7-230}
\begin{array}{l}
3a)~B^{[0]}=\beta~\Rightarrow~K^{[0]}=k^{[0]}=k~,~K_{x}^{[0]}=k_{x0}^{[0]}=0~,~K_{z}^{[0]}=k~,\\
3b)~B^{[1]}=\beta~\Rightarrow~K^{[1]}=k~,~K_{z}^{[1]}=k~,\\
3c)~A^{[0]+}=a^{[0]+}~,\\
3d)~M^{[0]}=\mu^{[0]}~,\\
3e)~M^{[1]}=\mu^{[1]}\frac{w}{d}\\
3f)~H \text{ solution ($\ne~h$) of } \cos\left(K_{z}^{[1]}H\right)+\frac{M^{[1]}K_{z}^{[1]}}{iM^{[0]}K_{z}^{[0]}}\sin\left(K_{z}^{[1]}H\right)=\\
~~~\cos\left(kh\right)+\frac{w}{id}\sin\left(kh\right)~.\\
\end{array}
\end{equation}
Note that now only {\it 3a)-3e)} constitute explicit solutions for five of the layer problem parameters whereas the obtention of $H$ requires solving a nonlinear equation for each frequency and the solution of this equation is not unique.
\subsubsection{Further comments on consequences of the $M=0$ approximation of grating response}\label{comm}
If, for a reason to be evoked further on, one chooses one of the three solutions as a means of identifying some or all of the effective medium parameters (i.e., those denoted by upper-case letters), then he should be aware of the fact that these choices all derive from a $M=0$ approximation of the uneven boundary response, which, as shown previously in sect. \ref{Meq0}, is only valid at very low frequencies and  cannot account, by any means,  for resonant behavior of the uneven boundary.
\subsubsection{Numerical results for the effective medium response derived from the $M=0$ approximation of the uneven boundary response as well as from the effective shear modulus $M=\mu w/d$}\label{resMe0}
\begin{figure}[ht]
\begin{center}
\includegraphics[width=0.75\textwidth]{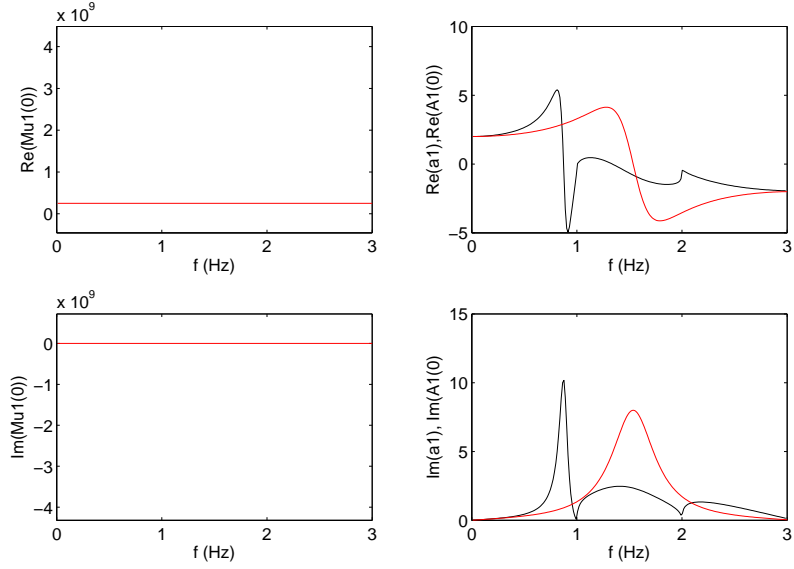}
\caption{The upper and lower left-hand panels represent the real and imaginary parts respectively of $M^{[1]}$ obtained, via $1a-1f$, from the $M=0$ approximation of  $a_{0}^{[1]}(f)$. In the right hand panels: the black curves represent the reference spectra $a_{0}^{[1]}(f)$ and the red curves the approximate spectra $a_{0}^{[1](0)}(f)=A^{[1]}$. The upper (lower) panels correspond to the real (imaginary) parts of these functions.  Case $w=430~m$,  $h=280~m$.}
\label{resMe0-01}
\end{center}
\end{figure}

\begin{figure}[ptb]
\begin{center}
\includegraphics[width=0.75\textwidth]{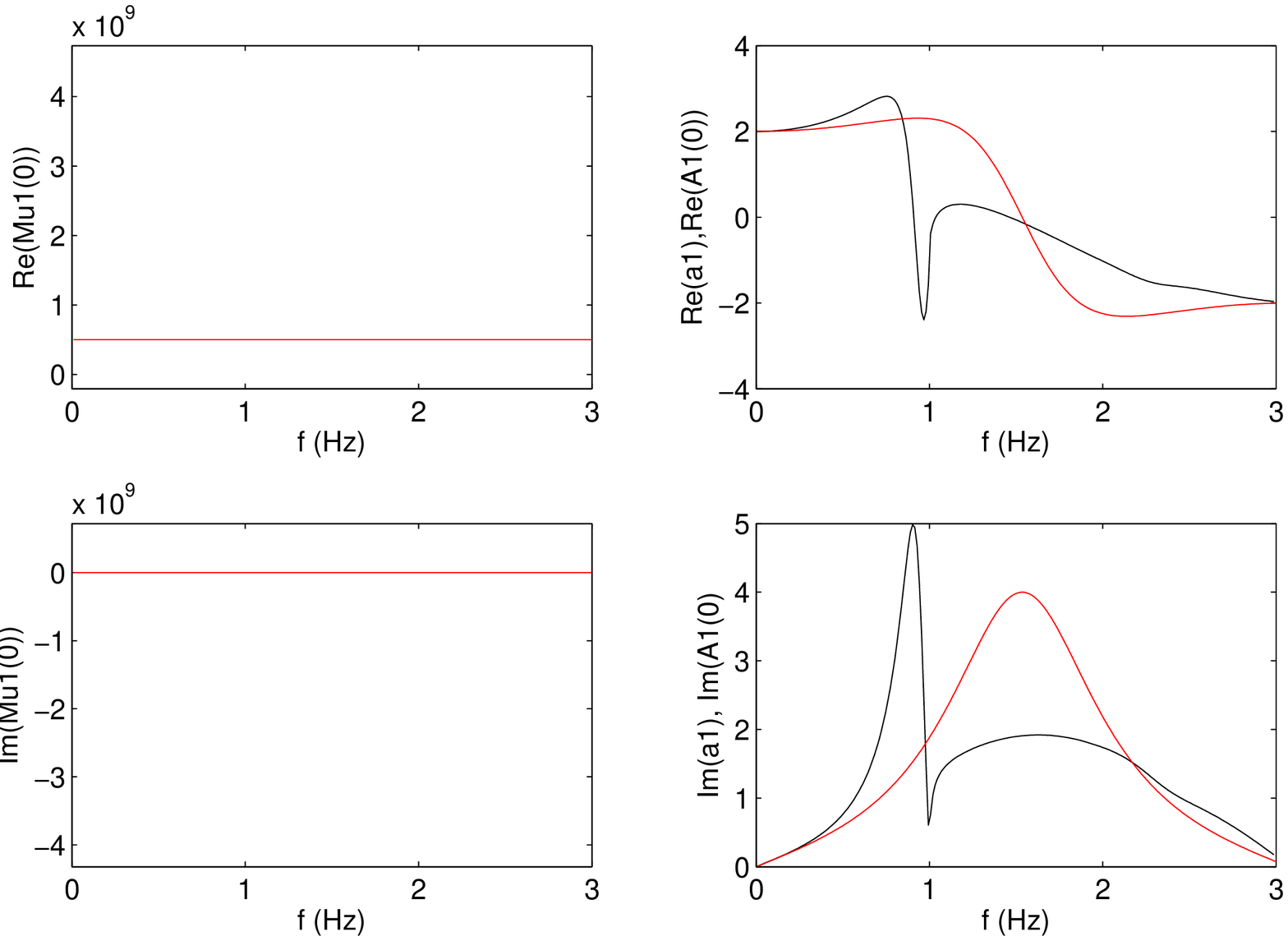}
\caption{Same as fig. \ref{resMe0-01} except that $w=860~m$,  $h=280~m$.}
\label{resMe0-02}
\end{center}
\end{figure}

\begin{figure}[ptb]
\begin{center}
\includegraphics[width=0.75\textwidth]{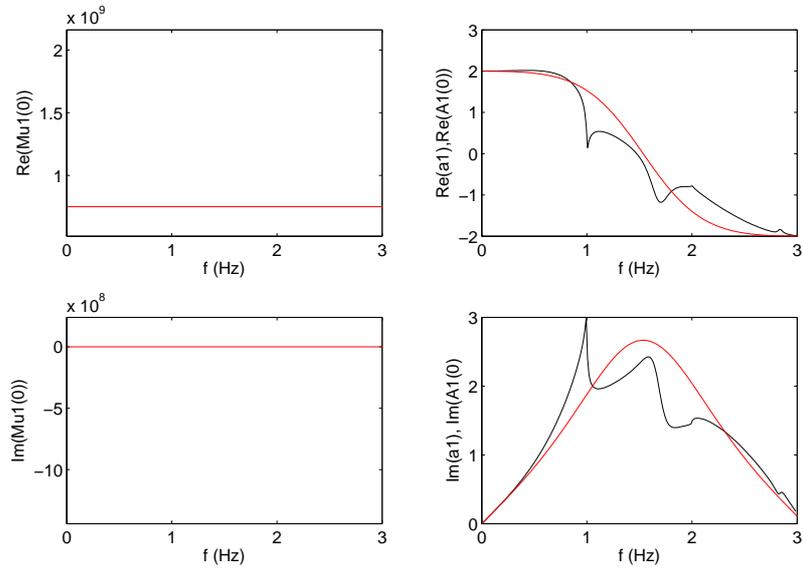}
\caption{Same as fig. \ref{resMe0-01} except that $w=1290~m$,  $h=280~m$.}
\label{resMe0-03}
\end{center}
\end{figure}

Figs. \ref{resMe0-01}-\ref{resMe0-03} show how the $M=0$ approximate response spectra evolve with $w$. In all these figures, $\theta^{i}=0^{\circ}$, $d=1720~m$, $\beta^{[0]}=17200~ms^{-1}$, $\mu^{[0]}=1\times 10^{9}~Pa$,  $\beta^{[1]}=1720~ms^{-1}$, $\mu^{[1]}=1\times 10^{9}~Pa$ and the reference solutions blue curves are obtained for $M=N=25$.
\clearpage
\newpage
We observe in these figures that, as expected, the effective layer approximation of response obtained from $M^{[1]}=\mu w/d$ does not account for the resonant behavior in the neighborhood of $f=1~Hz$.
\section{The $M=1$ approximation of uneven boundary response}
Previous graphs (in sects. \ref{numMredfw}-\ref{numMredfh}) show that there exists a similarity between the reference uneven boundary response and the $M=1$ approximation of this  response even in the neighborhood of what appear to be resonances. We now try to exploit  this numerical similarity by showing how it arises in theoretical manner. This will lead to the notion of an effective medium capable of accounting for a resonance associated with the excitation of a Love mode and to another resonant-like feature which we later qualify 'pseudo-resonance'.

The $M=1$ linear system that must be solved is
\begin{equation}\label{11-010}
\sum_{m=0}^{1}X_{lm}^{(1)}Y_{m}^{(1)}=Z_{l}~;~l=0,1~,
\end{equation}
wherein
\begin{equation}\label{11-020}
X_{lm}^{(1)}=\delta_{lm}\cos\big(k_{zm}^{[1]}h\big)+g\epsilon_{l}k_{zm}^{[1]}\sin\big(k_{zm}^{[1]}h\big)\Sigma_{lm}^{(1)}~~,~~
g=\frac{w}{4id}~~,~~\Sigma_{lm}^{(1)}=\sum_{n=-1}^{1}\frac{1}{k_{zn}^{[0]}}E_{nl}^{+}E_{nm}^{-}~,
\end{equation}
\begin{equation}\label{11-030}
Y_{m}^{(1)}=a_{m}^{[1](1)}~,~~Z_{l}=a^{[0]+}\epsilon_{l}E_{0l}^{+}~~,~~
E_{nm}^{\pm}=i^{m}\text{sinc}\Big[\big(\pm k_{xn}^{[0]}+k_{xm}^{[1]}\big)\frac{w}{2}\Big]+
i^{-m}\text{sinc}\Big[\big(\pm k_{xn}^{[0]}-k_{xm}^{[1]}\big)\frac{w}{2}\Big]
~,
\end{equation}
from which it follows that:
\begin{equation}\label{11-040}
Y_{0}^{(1)}=\frac{N_{0}^{(1)}}{D^{(1)}}=\frac{Z_{0}X_{11}^{(1)}-Z_{1}X_{01}^{(1)}}{X_{00}^{(1)}X_{11}^{(1)}-X_{10}^{(1)}X_{01}^{(1)}}
~~,~~Y_{1}^{(1)}=\frac{N_{1}^{(1)}}{D^{(1)}}=
\frac{Z_{1}X_{00}^{(1)}-Z_{0}X_{10}^{(1)}}{X_{00}^{(1)}X_{11}^{(1)}-X_{10}^{(1)}X_{01}^{(1)}}
~.
\end{equation}
Now, as previously, we assume normal-incidence ($\theta^{i}=0^{\circ}$) incidence. which entails $k_{xn}^{[0]}=2n\pi/d=k_{x-n}^{[0]}$ and $k_{x0}^{[0]}=0$ whence
\begin{equation}\label{11-070}
E_{n0}^{\pm}=2\text{sinc}\Big[\frac{n\pi w}{d}\Big]~~,~~E_{00}^{\pm}=2
~,
\end{equation}
\begin{equation}\label{11-080}
E_{01}^{\pm}=0~~,~~E_{-n1}^{\pm}=-E_{n1}^{\pm}
~.
\end{equation}
Consequently:
\begin{equation}\label{11-090}
\Sigma_{00}^{[1)}=\frac{4}{k_{z0}^{[0]}}+\frac{8}{k_{z1}^{[0]}}\Big[\text{sinc}\big(\frac{\pi w}{d}\big)\Big]^{2}~~,~~\Sigma_{01}^{[1)}=\Sigma_{10}^{[1)}=0
~,
\end{equation}
so that
\begin{equation}\label{11-100}
D^{(1)}=\left[\cos\big(k_{z0}^{[1]}h\big)+gk_{z0}^{[1]}\sin\big(k_{z0}^{[1]}h\big)\Sigma_{00}^{(1)}\right]
\left[\cos\big(k_{z1}^{[1]}h\big)+2gk_{z1}^{[1]}\sin\big(k_{z1}^{[1]}h\big)\Sigma_{11}^{(1)}\right]~,
\end{equation}
\begin{equation}\label{11-110}
N_{0}^{(1)}=a^{0+}2\left[\cos\big(k_{z1}^{[1]}h\big)+2gk_{z1}^{[1]}\sin\big(k_{z1}^{[1]}h\big)\Sigma_{11}^{(1)}\right]~,
\end{equation}
\begin{equation}\label{11-120}
N_{1}^{(1)}=0
\end{equation}
whence
\begin{equation}\label{11-130}
Y_{0}^{(1)}=\frac{2a^{0+}}{\cos\big(k_{z0}^{[1]}h\big)+gk_{z0}^{[1]}\sin\big(k_{z0}^{[1]}h\big)\left[\frac{4}{k_{z0}^{[0]}}+\frac{8}{k_{z1}^{[0]}}\Big[\text{sinc}\big(\frac{\pi w}{d}\big)\Big]^{2}\right]}~~,~~Y_{1}^{(1)}=0~.
\end{equation}
%
\section{Relation of the approximate $M=1$ and $M=0$ solutions of the uneven boundary problem}
Since, by definition, $a_{1}^{[1](0)}=0$, it follows from (\ref{11-130}) that under all circumstances
\begin{equation}\label{11-135}
a_{1}^{[1](1)}=a_{1}^{[1](0)}=0~.
\end{equation}
Taking account of the fact that we are in normal incidence, (\ref{11-130}) becomes
\begin{equation}\label{11-140}
a_{0}^{[1](1)}=\frac{2a^{0+}}{\cos(kh)+
\frac{w}{id}\sin(kh)\left[1+\frac{2k}{k_{z1}^{[0]}}\Big[\text{sinc}\big(\frac{\pi w}{d}\big)\Big]^{2}\right]}~~,~~a_{1}^{[1](1)}=0~.
\end{equation}
Since sinc$(\pi)=0$, we find, by comparison with (\ref{6-030}), that
\begin{equation}\label{11-145}
\lim_{w/d\rightarrow 1}a_{0}^{[1](1)}=a_{0}^{[1](0)}=\frac{2a^{0+}}{\cos(kh)+\frac{w}{id}\sin(kh)}~.
\end{equation}
Finally, since $\sin(0)=0$, we  find, again by comparison of (\ref{11-140}) with \ref{6-030}), that
\begin{equation}\label{11-150}
\lim_{kh\rightarrow 0}a_{0}^{[1](1)}=a_{0}^{[1](0)}=2a^{0+}~.
\end{equation}
Thus, the $M=0$ solution can be expected to yield an approximation of the uneven boundary response that is all the better the closer is $w/d$ to 1 and/or $kh$ is closer to 0. This, of course, is what was observed in the previous numerical results concerning the uneven boundary response.
\section{Theoretical explanation, via the $M=1$ approximation of uneven boundary response, of the occurrence of a resonance near, and at a frequency lower than, $f=f_{WF}$ }\label{lmreson}
We can write (\ref{11-130}) as
\begin{equation}\label{11-160}
a_{0}^{[1](1)}=\frac{2a^{0+}}
{D_{1}+D_{2}}
~,
\end{equation}
wherein
\begin{equation}\label{11-170}
D_{1}=\cos(k_{z0}^{[1]}h\big)+
\frac{k_{z0}^{[1]}}{k_{z1}^{[0]}}
\frac{2w}{id}\Big[\text{sinc}\big(\frac{\pi w}{d}\big)\Big]^{2}
\sin\big(k_{z0}^{[1]}h\big)~~,~~
D_{2}=\frac{k_{z0}^{[1]}}{k_{z0}^{[0]}}\frac{w}{id}\sin\big(k_{z0}^{[1]}h\big)
~.
\end{equation}
or, more explicitly, on account of the fact that $\theta^{i}=0$
\begin{equation}\label{11-175}
D_{1}=\cos(kh)+
\frac{k}{k_{z1}^{[0]}}
\frac{2w}{id}\Big[\text{sinc}\big(\frac{\pi w}{d}\big)\Big]^{2}
\sin(kh)~~,~~
D_{2}=\frac{w}{id}\sin(kh)
~.
\end{equation}

A resonance is known to occur at $f=f_{R}$ in a function of $f$ such as $N(f)/D(f)$ when $D(f_{R})$ (but not $N(f_{R})$) is close to zero. We showed previously in sect. \ref{Meq0}  that $a_{0}^{[1](0)}$ does not exhibit resonant behavior at any frequency because the denominator  $D_{0}$ of $a_{0}^{[1](0)}$ is never close to zero. On the contrary, the fact that  $a_{0}^{[1](1)}$ has been found numerically to exhibit resonant behavior in the neighborhood of $f=1~Hz$ (which is equal to the Wood-Fano frequency $f_{WF}$ discussed further on) would seem to require one of the following conditions at $f_{R}$:\\\\
 (i) $D_{1}$ and $D_{2}$ are both real and of opposite signs,\\
 (ii) $D_{1}$ and $D_{2}$ are both imaginary and of opposite signs,\\
 (iii) $D_{1}$ real but close to or equal to zero and $D_{2}$ imaginary but small.\\\\
 Recall that we assumed at the outset  that the medium  in $\mathcal{L}$ is lossless, this meaning that $k$ is real. It follows that $k_{z0}^{[1]}=k$,  $\cos\big(k_{z0}^{[1]}h\big)$ and $\sin\big(k_{z0}^{[1]}h\big)$ are all real. Similarly, $k_{z0}^{[0]}=k$ is real which means (because $w/d$ is real) that $D_{2}$ is imaginary at all frequencies, this entailing that the case (i) is not possible. Recall that $D_{1}=D_{11}+D_{12}$, with $D_{11}$ the $\cos$ term and $D_{12}$ the $\sin$ term. Under the previous assumptions, $D_{11}$ is real which seems to exclude case (ii) unless the $\cos$ term equals 0 so that we must compare $D_{12}$ to $D_{2}$ to see if they are both imaginary and of opposite signs. It is true that $D_{2}$ is imaginary and $D_{12}$ might be imaginary, but they are not of opposite signs so case (ii) is impossible. Let us now examine whether case (iii) is possible. $D_{1}$ real means that $D_{12}=\frac{k}{k_{z1}^{[0]}}
\frac{2w}{id}\Big[\text{sinc}\big(\frac{\pi w}{d}\big)\Big]^{2}
\sin(kh)$ must be real which (under the previous assumptions)  can only occur when $k_{z1}^{[0]}=\sqrt{k^{2}-\big(2\pi/d\big)^{2}}$ is imaginary. This can {\it possibly} occur for frequencies such that $k^{[0]}<2\pi/d$ (i.e., $f<f_{WF}=\beta/d$) and $D_{1}$ will be able to equal zero and a resonance to {\it certainly} occur when
\begin{equation}\label{11-200}
D_{1}=\cos(kh)-
\frac{k}{\sqrt{\big(2\pi/d\big)^{2}}-k^{2}}
\frac{2w}{d}\Big[\text{sinc}\big(\frac{\pi w}{d}\big)\Big]^{2}
\sin(kh)=0
~,
\end{equation}
the resonance being all the more pronounced, the smaller is the modulus of $D_{2}$. Eq. (\ref{11-200}) turns out to be nothing else but the dispersion relation for what appear to be Love modes (LM) \cite{ejp57,ke01,wi88} (actually somewhat different from the Love modes of  a homogeneous layer/homogeneous half space configuration because of the structure factor of the uneven boundary). Note that this dispersion relation involves all the geometrical and constitutive properties of the uneven boundary configuration. Thus, case (iii) can be realized which shows theoretically  under what circumstances a genuine resonance be generated in the response ($a_{0}^{[1](1)}$) of the uneven boundary. Note that (\ref{11-200}) can be realized only if $D_{11}$  is positive since $D_{12}$ is negative, so that a necessary condition for a LM to be excited is that
\begin{equation}\label{11-201}
(2n+1)\frac{\pi}{2}>kh>2n\pi~;~n=0,1,2,...
~,
\end{equation}
or, for the low frequencies of interest herein
\begin{equation}\label{11-202}
\frac{\pi}{2}>kh>0~\Rightarrow~\frac{\beta}{4h}>f>0
~.
\end{equation}
%
\section{Theoretical explanation, via the $M=1$ approximation of uneven boundary response, of the occurrence of what appears to be another type of resonance near, and at a frequency higher than $f=f_{WF}$}\label{fbswreson}
We can write (\ref{11-130}) as
\begin{equation}\label{11-203}
a_{0}^{[1](1)}=\frac{2a^{0+}}
{D_{3}+D_{4}}
~,
\end{equation}
wherein
\begin{equation}\label{11-204}
D_{3}=\cos(k_{z0}^{[1]}h\big)~~,~~
D_{4}=
\frac{w}{id}\left[\frac{k_{z0}^{[1]}}{k_{z0}^{[0]}}+2\frac{k_{z0}^{[1]}}{k_{z1}^{[0]}}\Big[\text{sinc}\big(\frac{\pi w}{d}\big)\Big]^{2}
\right]\sin\big(k_{z0}^{[1]}h\big)
~,
\end{equation}
or, more explicitly, on account of the fact that $\theta^{i}=0$
\begin{equation}\label{11-206}
D_{3}=\cos(kh)~~,~~
D_{4}=\frac{w}{id}\left[1+2\frac{k}{k_{z1}^{[0]}}\Big[\text{sinc}\big(\frac{\pi w}{d}\big)\Big]^{2}
\right]\sin(kh)
~.
\end{equation}
The expression
\begin{equation}\label{11-208}
D_{3}=\cos(kh)=0
~.
\end{equation}
is the 'dispersion relation'  of the so-called fixed-base shear wall  modes (FBSWM) \cite{tr72} so that one might expect a sort of resonance to occur at the frequencies
\begin{equation}\label{11-210}
f=\frac{(2n+1)\beta}{4h}~;~n=0,1,2,...
~.
\end{equation}
Actually, since (\ref{11-208}) is not a true dispersion relation of the uneven boundary configuration due to the fact that it does not incorporate all the parameters (notably $w$ and $d$) of the latter, the FBSWM  are not true modes of the uneven boundary configuration. Thus, we term these 'pseudo-modes' and note that the response is not maximal at the FBSWM 'resonance' frequencies but rather at the frequencies for which $\|D_{3}+D_{4}\|$ is minimal. This will be illustrated in the numerical results given further on.
\section{Relation of the approximate $M=1$ solution of the uneven boundary problem to the exact solution of the homogeneous layer problem as a means of obtaining an effective medium representation of the uneven boundary}
The exact solution for the homogeneous layer over homogeneous half space problem was given in (\ref{7-190}), which for normal incidence (i.e., $K_{z}^{[1]}=K^{[1]}$  and $K_{z}^{[0]}=K^{[0]}$  becomes
\begin{equation}\label{11-210}
A^{[1]}=A^{[0]+}\left[\frac{2}
{\cos\left(K^{[1]}H\right)+\frac{M^{[1]}K^{[1]}}{iM^{[0]}K^{[0]}}\sin\left(K^{[1]}H\right)}\right]~.
\end{equation}
which we now compare to the expression of $a_{0}^{[1](1)}$ in (\ref{11-140}).

We assume, similarly to what was done in sect. \ref{first} (recalling  $\Theta^{i}=\theta^{i}=0$), that:
\begin{equation}\label{11-220}
\begin{array}{l}
1a)~B^{[0]}=\beta~\Rightarrow~K^{[0]}=k~,~K_{x}^{[0]}=k_{x0}^{[0]}=0~,~K_{z}^{[0]}=k_{z0}^{[0]}=k~,\\
1b)~B^{[1]}=\beta~\Rightarrow~K^{[1]}=k~,~K_{z}^{[1]}=k_{z0}^{[1]}=k~,\\
1c)~A^{[0]+}=a^{[0]+}~,\\
1d)~H=h~,\\
1e)~M^{[0]}=\mu~.\\
\end{array}
\end{equation}
Note that we no longer make the assumption $1f$ because the latter is a direct consequence of the $M=0$ approximation of the uneven boundary response whereas now we are relying on the $M=1$ approximation of this response. Thus, integrating $1a)-1e$ into (\ref{11-210}) transforms the latter into
\begin{equation}\label{11-230}
A^{[1]}=a^{[0]+}\left[\frac{2}
{\cos(kh)+\frac{M^{[1]}}{i}\sin(kh)}\right]~,
\end{equation}
which compared to (\ref{11-140}), i.e.,
\begin{equation}\label{11-240}
a_{0}^{[1](1)}=A^{[1]}~,
\end{equation}
translates to
\begin{equation}\label{11-250}
M^{[1]}=\mu \frac{w}{d}\left[1+2\frac{k}{k_{z1}^{[0]}}\Big[\text{sinc}\Big(\frac{\pi w}{d}\Big)\Big]^{2}\right]~,
\end{equation}
which is nothing other than the shear modulus of the effective layer in the sense of the equality relation (\ref{11-240}), the latter being a short-hand expression of the equality of the response field on  the uppermost segments of the uneven boundary to the response field on the topmost boundary of the layer, since these response fields are
\begin{equation}\label{11-260}
u^{[1](1)}(x,h,\omega)=a_{0}^{[1](1)}(\omega)~~,~~U^{[1]}(x,H=h,\omega)=A^{[1]}(\omega)~~;     nd-\frac{w}{2}\le x\le nd+\frac{w}{2}~,n=0,\pm 1,....
\end{equation}
\clearpage
\newpage
\subsection{Numerical results for the effective medium response derived from the $M=1$ approximation of the uneven boundary response}\label{numMeq1}
\begin{figure}[ht]
\begin{center}
\includegraphics[width=0.75\textwidth]{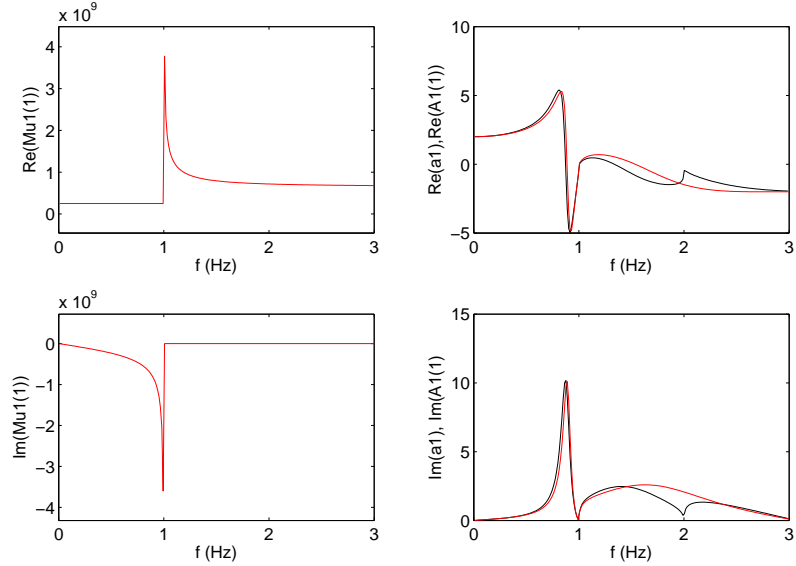}
\caption{The upper and lower left-hand panels represent the real and imaginary parts respectively of $M^{[1]}$ obtained, via $1a-1e$ and (\ref{11-250}), from the $M=1$ approximation of  $a_{0}^{[1]}(f)$. In the right hand panels: the black curves represent the reference spectra $a_{0}^{[1]}(f)$ and the red curves the approximate spectra $a_{0}^{[1](1)}(f)$. The upper (lower) panels correspond to the real (imaginary) parts of these functions.  Case $w=430~m$,  $h=280~m$.}
\label{numMeq1-01}
\end{center}
\end{figure}

\begin{figure}[ptb]
\begin{center}
\includegraphics[width=0.75\textwidth]{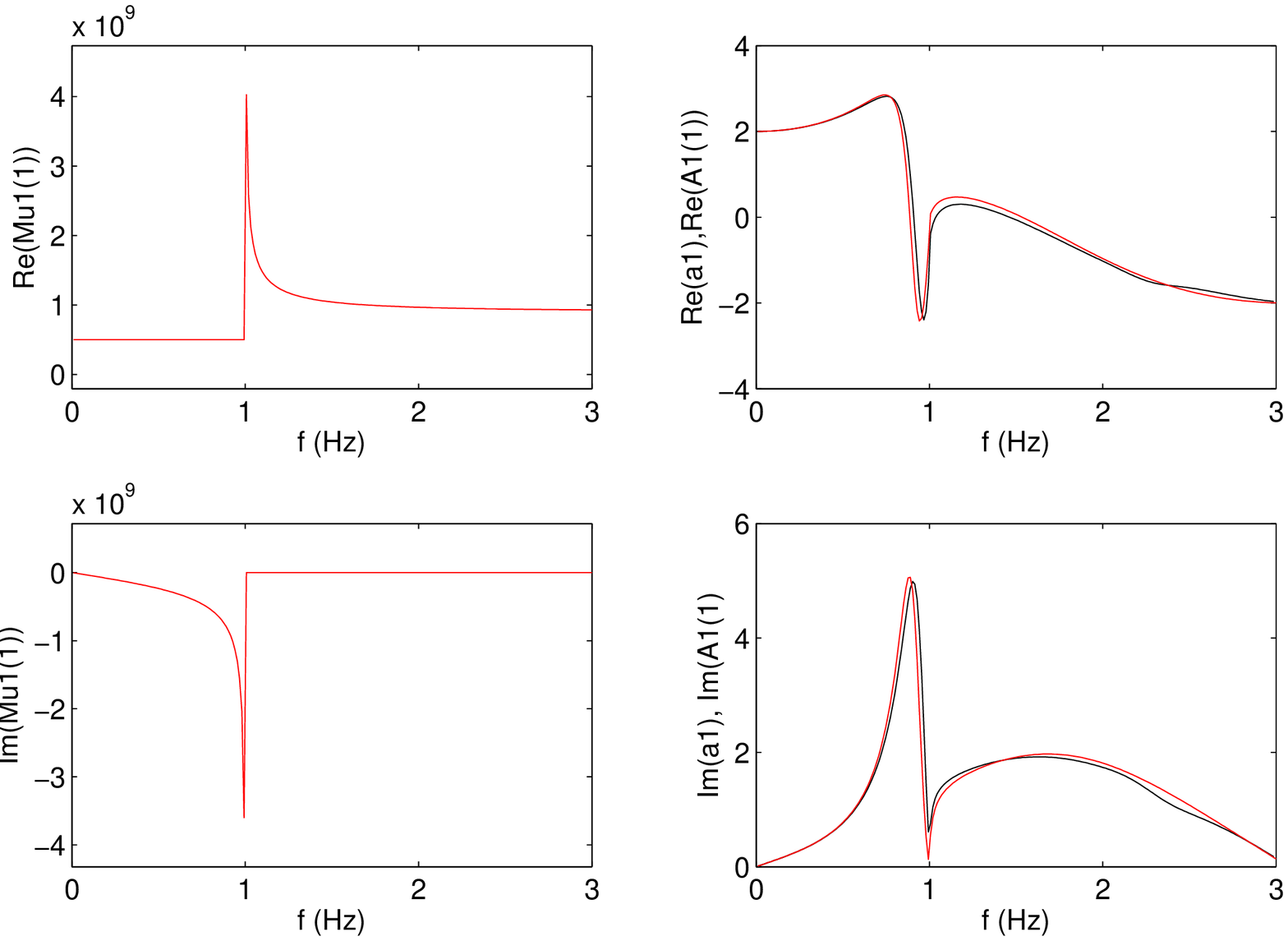}
\caption{Same as fig. \ref{numMeq1-01} except that $w=860~m$,  $h=280~m$.}
\label{numMeq1-02}
\end{center}
\end{figure}

\begin{figure}[ptb]
\begin{center}
\includegraphics[width=0.75\textwidth]{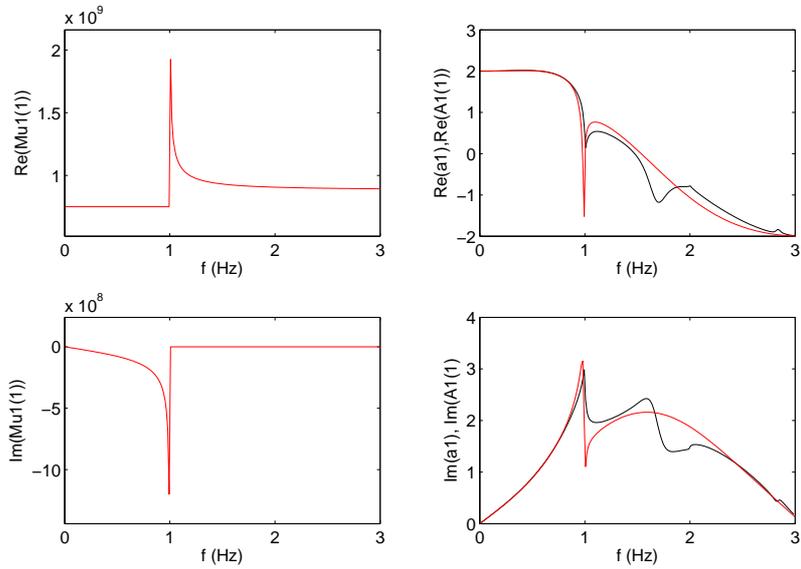}
\caption{Same as fig. \ref{numMeq1-01} except that $w=1290~m$,  $h=280~m$.}
\label{numMeq1-03}
\end{center}
\end{figure}

Figs. \ref{numMeq1-01}-\ref{numMeq1-03} show how the $M=1$ effective shear modulus and approximate response spectra evolve with $w$. In all these figures, $\theta^{i}=0^{\circ}$, $d=1720~m$, $\beta^{[0]}=17200~ms^{-1}$, $\mu^{[0]}=1\times 10^{9}~Pa$,  $\beta^{[1]}=1720~ms^{-1}$, $\mu^{[1]}=1\times 10^{9}~Pa$ and the reference solutions black curves are obtained for $M=N=25$.
\clearpage
\newpage
As expected, we now see that both the new effective shear modulus and the $M=1$ approximation enable the lowest-frequency resonances to be accounted for.
\section{Case studies to identify the causes of the main features of response}\label{casestudy}
By means of the results depicted in figs. \ref{casest-01}-\ref{casest-12} we shall attempt to identify the causes of the principal features of low-frequency response of the periodically-uneven boundary of interest in this study. Actually, we treat six cases: the first three devoted to variations of $w$ for large heights $h$ and the last three to variations of $w$ for small heights $h$. Each case is illustrated by two figures: the first  deals with the effective shear modulus and the corresponding $M=1$ approximation of response spectra, whereas the second deals with the the various terms $D_{1}-D_{4}$ in the denominators of the $M=1$ response as well as this response. The underlying idea is that the principal features of response are strongly-related to the occurrence of LM resonances or FBSWM pseudo-resonances.

A word is here in order about the these resonances and pseudo-resonances (in fact we shall deal only with the lowest-frequency specimens of each of these). Previous results showed that the frequency at which $k_{z1}^{[0]}$ is nil is critical in that it marks the frontier at which the $M=1$ effective shear modulus changes from complex values (for the smaller $f$) to  real value values (for the larger $f$). The frequency at this frontier is termed $f_{WF}$ in honor of Wood and Fano \cite{fa41} who showed, and tried to explain, the anomalous behavior of response of periodically-uneven boundaries in the immediate neighborhood of this frequency. The discussion on LM resonances showed that the lowest-frequency resonance of this type occurs for $f<f_{WF}$ and it can be shown that the lowest-frequency FBSWM pseudo-resonance  occurs for $f>f_{WF}$.
Moreover, since it was assumed that the medium underneath the unveven boundary is lossless, the term $k_{z1}^{[0]}$ vanishes at $f=f_{WF}$, so that $D$ blows up at this frequency whence $a_{0}^{[1](1)}(f_{WF})=0$ which means that the $M=1$ predicted response vanishes at the Wood-Fano frequency.
\subsection{Variation of $w$ for large $h$}\label{casestudylh}
\begin{figure}[ht]
\begin{center}
\includegraphics[width=0.75\textwidth]{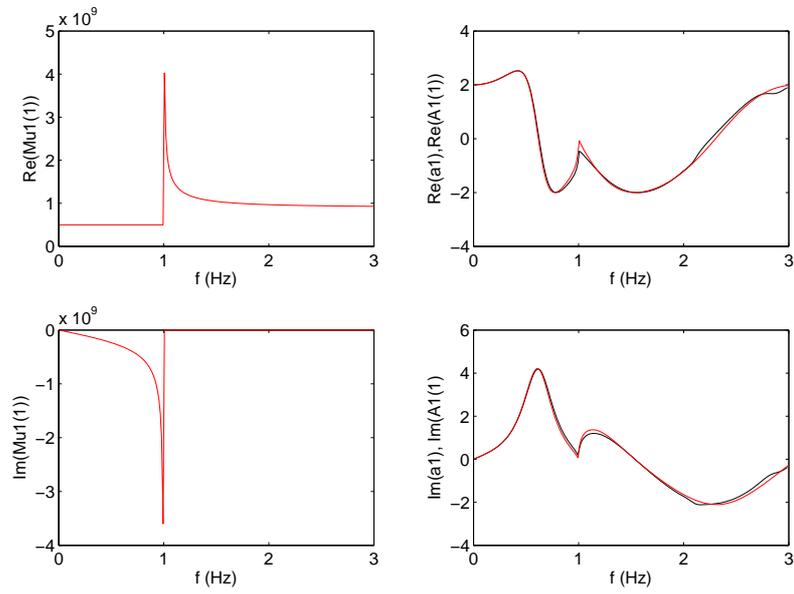}
\caption{The upper and lower left-hand panels represent the real and imaginary parts respectively of $M^{[1]}$ obtained, via $1a-1e$ and (\ref{11-250}), from the $M=1$ approximation of  $a_{0}^{[1]}(f)$. In the right hand panels: the black curves represent the reference spectra $a_{0}^{[1]}(f)$ and the red curves the approximate spectra $a_{0}^{[1](1)}(f)$. The upper (lower) panels correspond to the real (imaginary) parts of these functions.  Case $w=860~m$,  $h=560~m$,}
\label{casest-01}
\end{center}
\end{figure}

\begin{figure}[ptb]
\begin{center}
\includegraphics[width=0.75\textwidth]{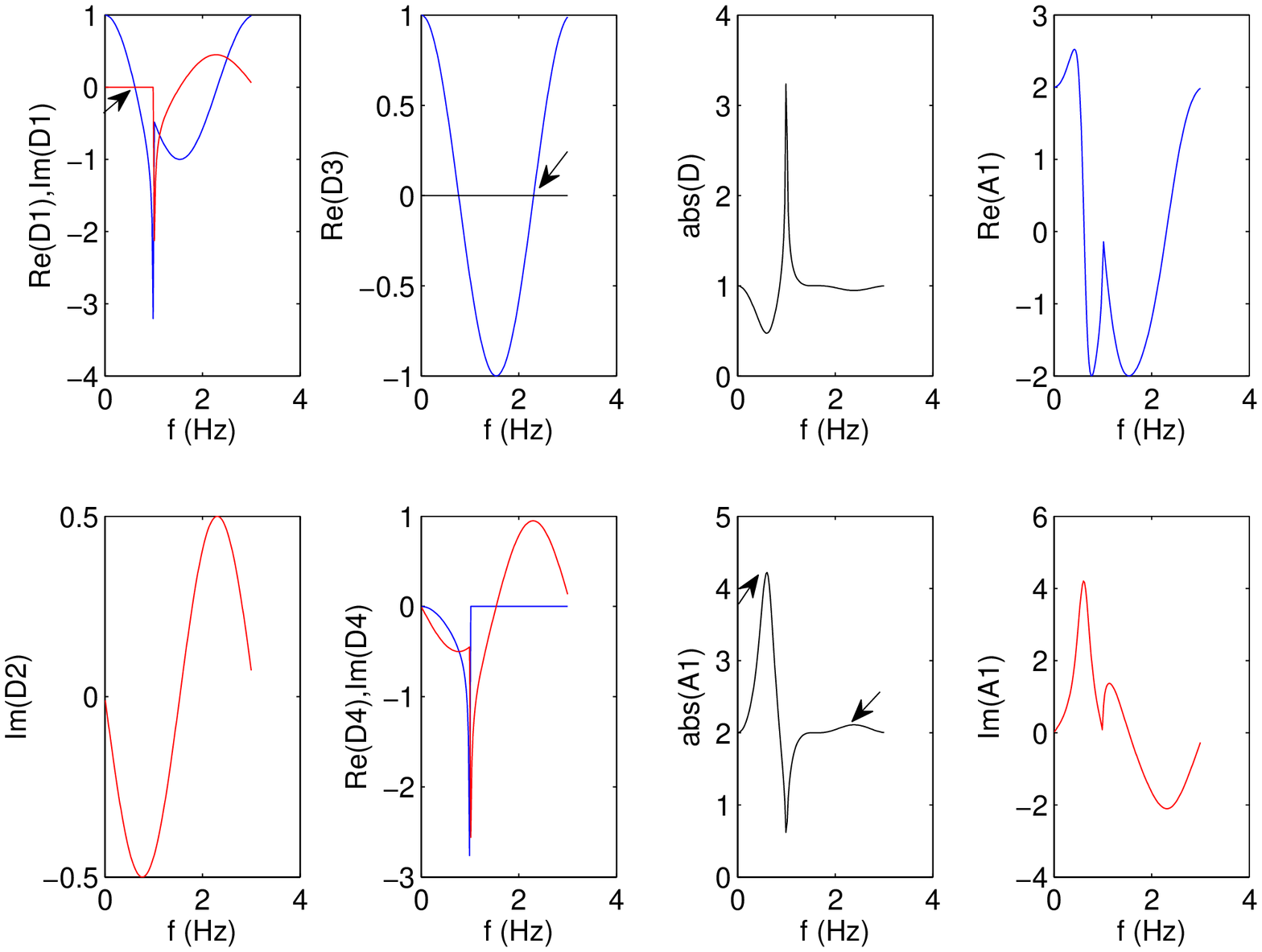}
\caption{The upper left-most panel depicts $\Re(D_{1}(f))$ in blue and  $\Im(D_{1}(f))$ in red. The second upper panel depicts $\Re(D_{3}(f))$ in blue and the zero level in black. The next upper panel depicts $\|D(f)\|$ and the right-most upper panel (an expanded version of which is given in the upper right-hand panel of the preceding figure) depicts $\Re(a_{0}^{[1](1)}(f))=\Re(A^{[1]}(f))$. The lower left-most panel depicts $\Im(D_{2}(f))$.  The second lower  panel depicts $\Re(D_{4}(f))$ in blue and  $\Im(D_{4}(f))$ in red. The next lower panel depicts $\|a_{0}^{[1](1)}(f)\|=\|A^{[1]}(f)\|$.  The right-most lower panel (an expanded version of which is given in the lower right-hand panel of the preceding figure) depicts $\Im(a_{0}^{[1](1)}(f))=\Im(A^{[1]}(f))$. Arrows pointing upwards denote locations of LM resonances and arrows pointing downwards denote locations of FBSWM resonances. Case $w=860~m$,  $h=560~m$.}
\label{casest-02}
\end{center}
\end{figure}

\begin{figure}[ptb]
\begin{center}
\includegraphics[width=0.75\textwidth]{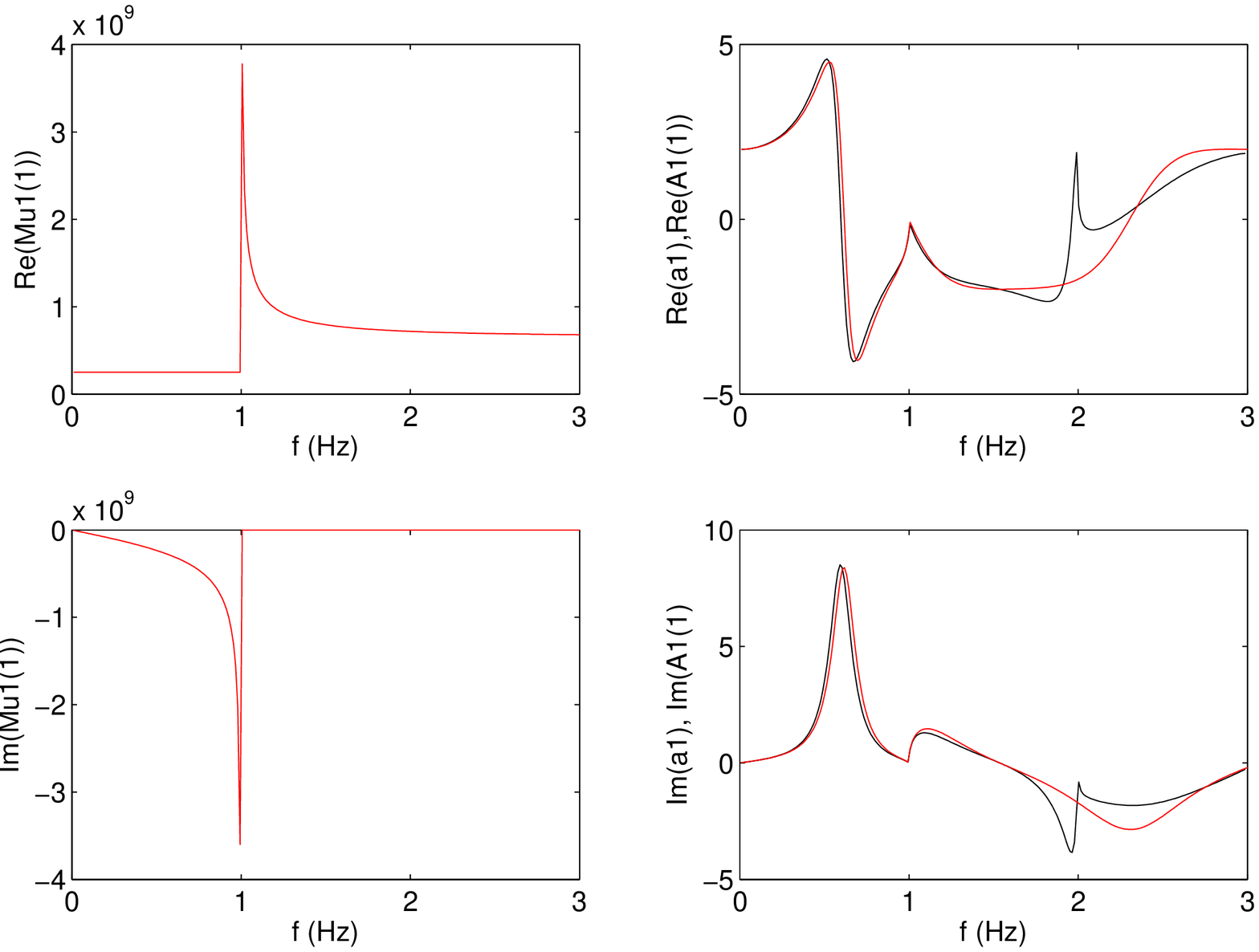}
\caption{Same as fig. \ref{casest-01} except that $w=430~m$,  $h=560~m$,}
\label{casest-03}
\end{center}
\end{figure}

\begin{figure}[ptb]
\begin{center}
\includegraphics[width=0.75\textwidth]{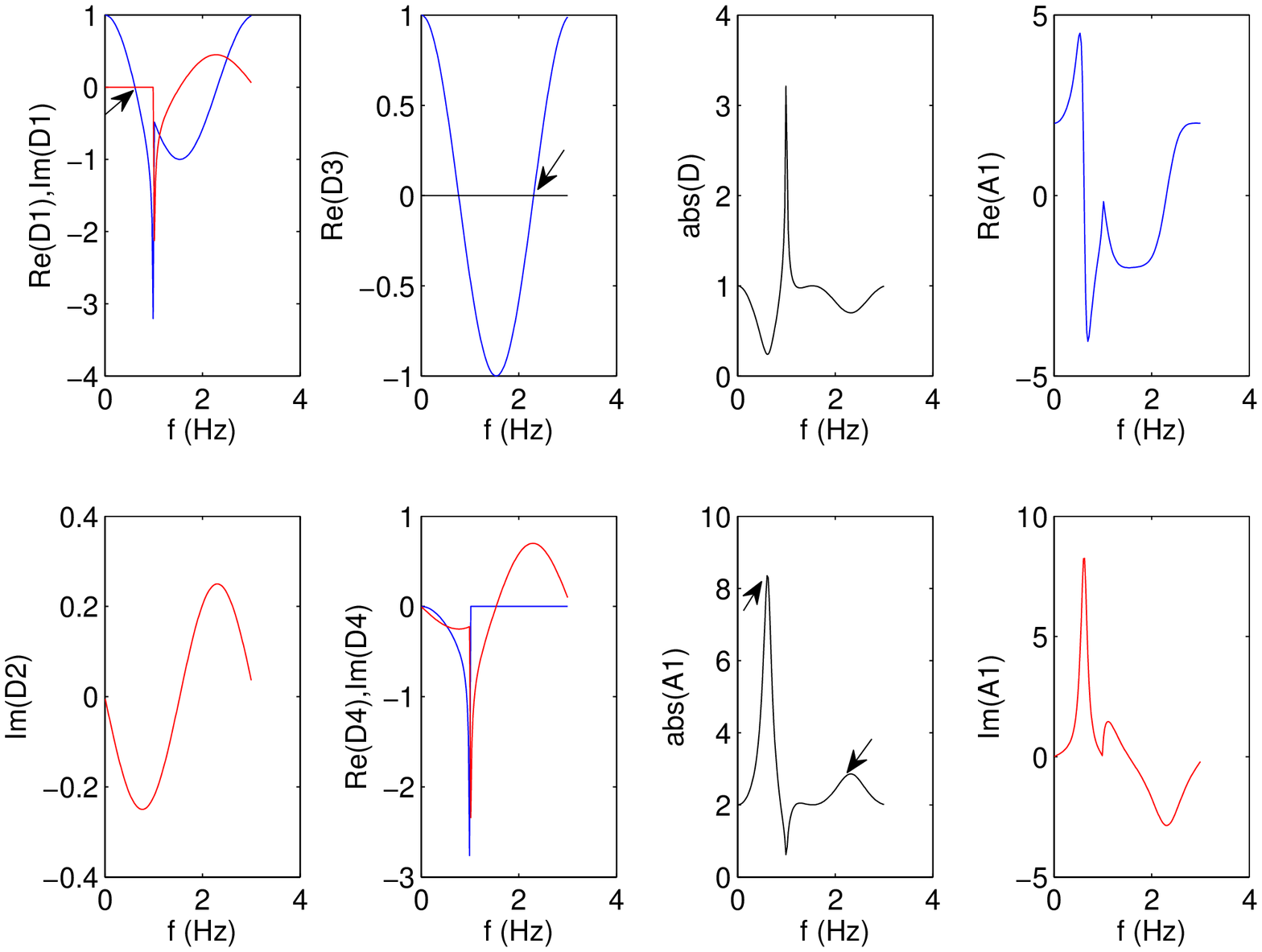}
\caption{Same as fig. \ref{casest-02} except that $w=430~m$,  $h=560~m$,}
\label{casest-04}
\end{center}
\end{figure}

\begin{figure}[ptb]
\begin{center}
\includegraphics[width=0.75\textwidth]{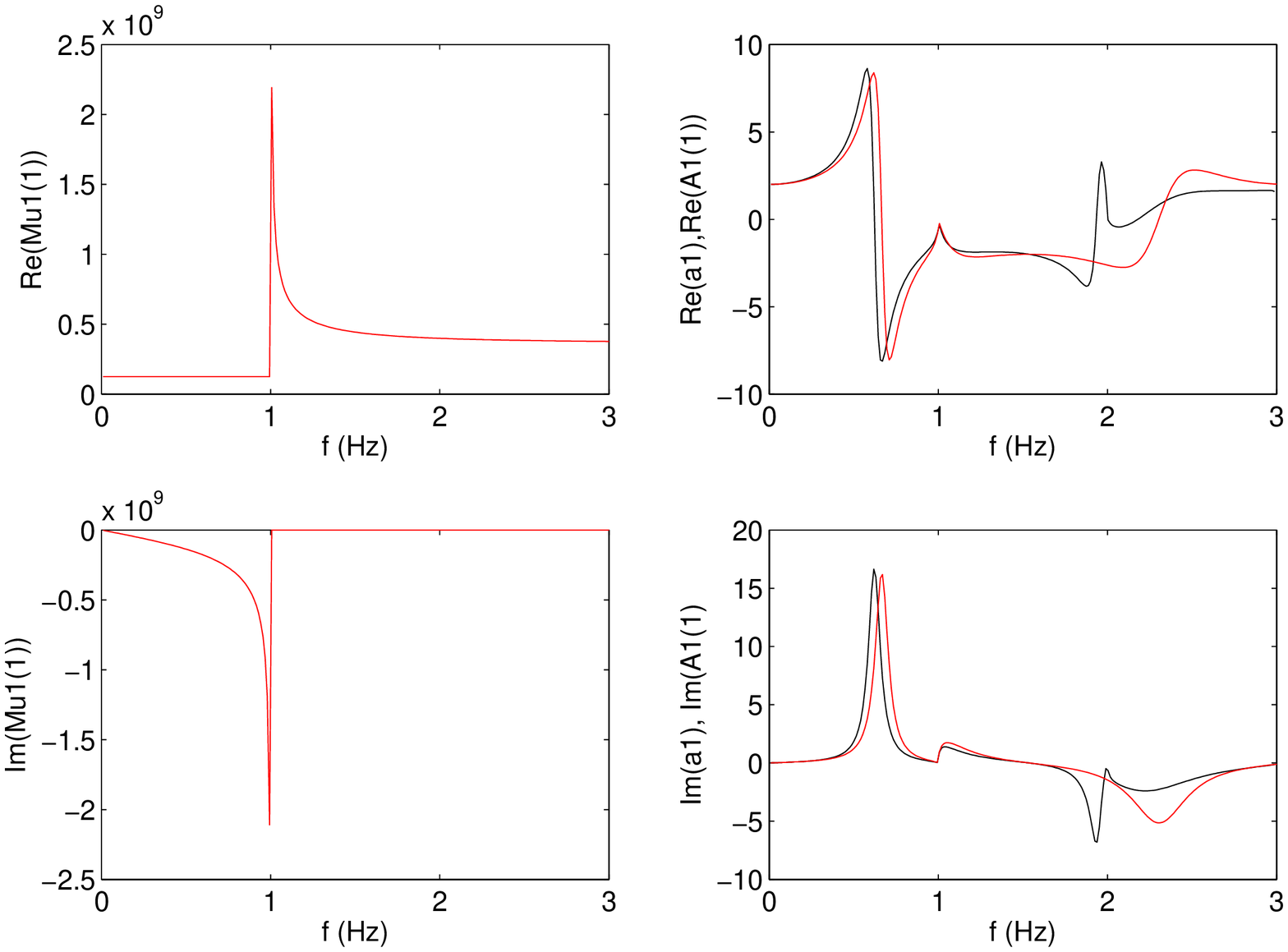}
\caption{Same as fig. \ref{casest-01} except that $w=215~m$,  $h=560~m$,}
\label{casest-05}
\end{center}
\end{figure}

\begin{figure}[ptb]
\begin{center}
\includegraphics[width=0.75\textwidth]{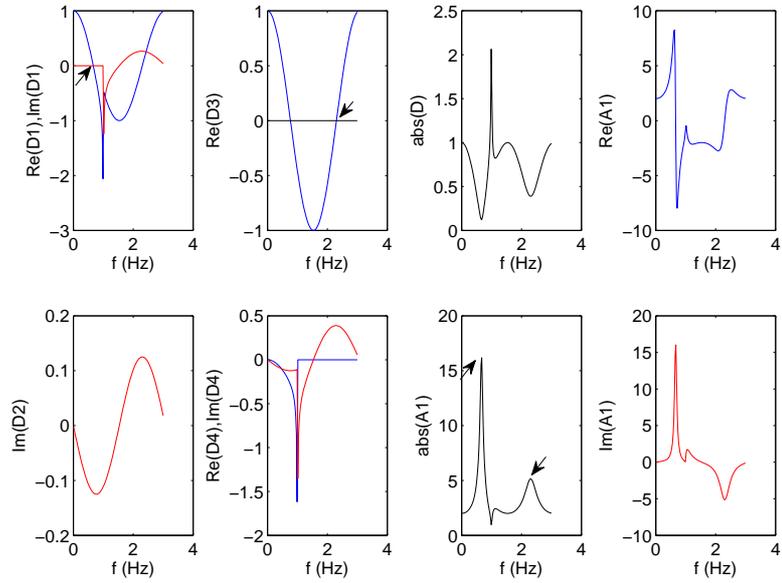}
\caption{Same as fig. \ref{casest-02} except that $w=215~m$,  $h=560~m$,}
\label{casest-06}
\end{center}
\end{figure}

Figs. \ref{casest-01}-\ref{casest-06} are relative to case studies of variation of $w$ for large-$h$ unveven boundaries. In all these figures, $\theta^{i}=0^{\circ}$, $d=1720~m$, $\beta^{[0]}=17200~ms^{-1}$, $\mu^{[0]}=1\times 10^{9}~Pa$,  $\beta^{[1]}=1720~ms^{-1}$, $\mu^{[1]}=1\times 10^{9}~Pa$ and the reference solutions  are obtained for $M=N=25$.
\clearpage
\newpage
Consider the prototypical response in the low-frequency interval $[0~Hz,1.5~Hz]$ depicted in figs. \ref{casest-03}-\ref{casest-04}.  This modulus of this response is characterized by two peaks, one, larger and narrower, to the left of $f_{WF}=1~Hz$, and the other, smaller and wider, to the right of $f_{WF}$. The left-hand peak is clearly due to the excitation of a LM as seen by the location (at the intersection of $\Re D_{1}(f)$ with the zero level) of the upward-pointing arrow in the upper left-most panel of fig. \ref{casest-04}. The righ-hand peak is clearly due to the excitation of a FBSWM as seen by the location (near the intersection of $\Re D_{3}(f)$ with the zero level) of the downward-pointing arrow in the upper second left-most panel of fig. \ref{casest-04}. The imaginary parts of $D$ at these locations (smaller for the LM than for the FBSWM) explain why the LM peak is higher than the FBSWM peak.

This set of figures also shows that the narrower  is the uneven feature (i.e., the smaller is $w$) of the boundary, the higher and narrower are both  the LM  and FBSWM peaks, although the LM peaks are always higher and narrower than the FBSWM peaks, this being probably due to the fact that the LM peak indicates the occurrence of a true resonance whereas the FBSWM  peak that of a pseudo-resonance.
\subsection{Variation of $w$ for small $h$}\label{casestudysh}
\begin{figure}[ht]
\begin{center}
\includegraphics[width=0.75\textwidth]{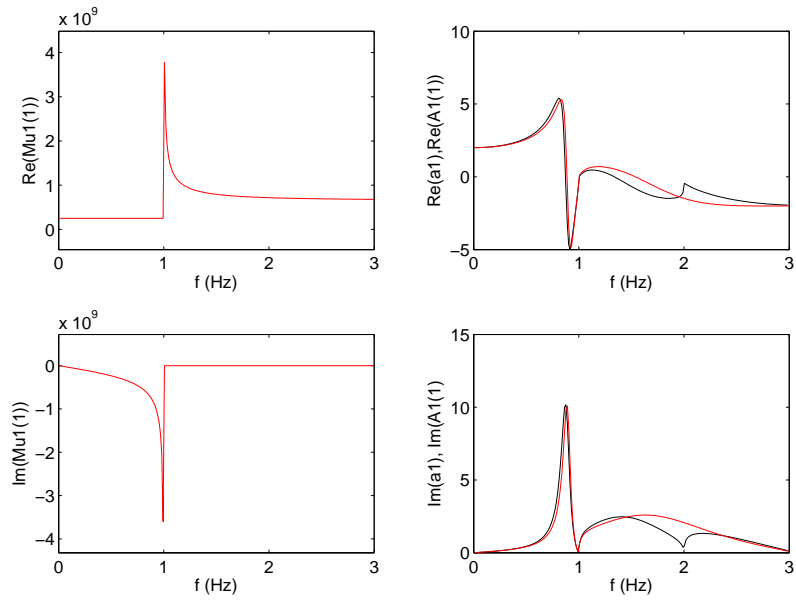}
\caption{The upper and lower left-hand panels represent the real and imaginary parts respectively of $M^{[1]}$ obtained, via $1a-1e$ and (\ref{11-250}), from the $M=1$ approximation of  $a_{0}^{[1]}(f)$. In the right hand panels: the black curves represent the reference spectra $a_{0}^{[1]}(f)$ and the red curves the approximate spectra $a_{0}^{[1](1)}(f)$. The upper (lower) panels correspond to the real (imaginary) parts of these functions.  Case $w=430~m$,  $h=280~m$,}
\label{casest-07}
\end{center}
\end{figure}

\begin{figure}[ptb]
\begin{center}
\includegraphics[width=0.75\textwidth]{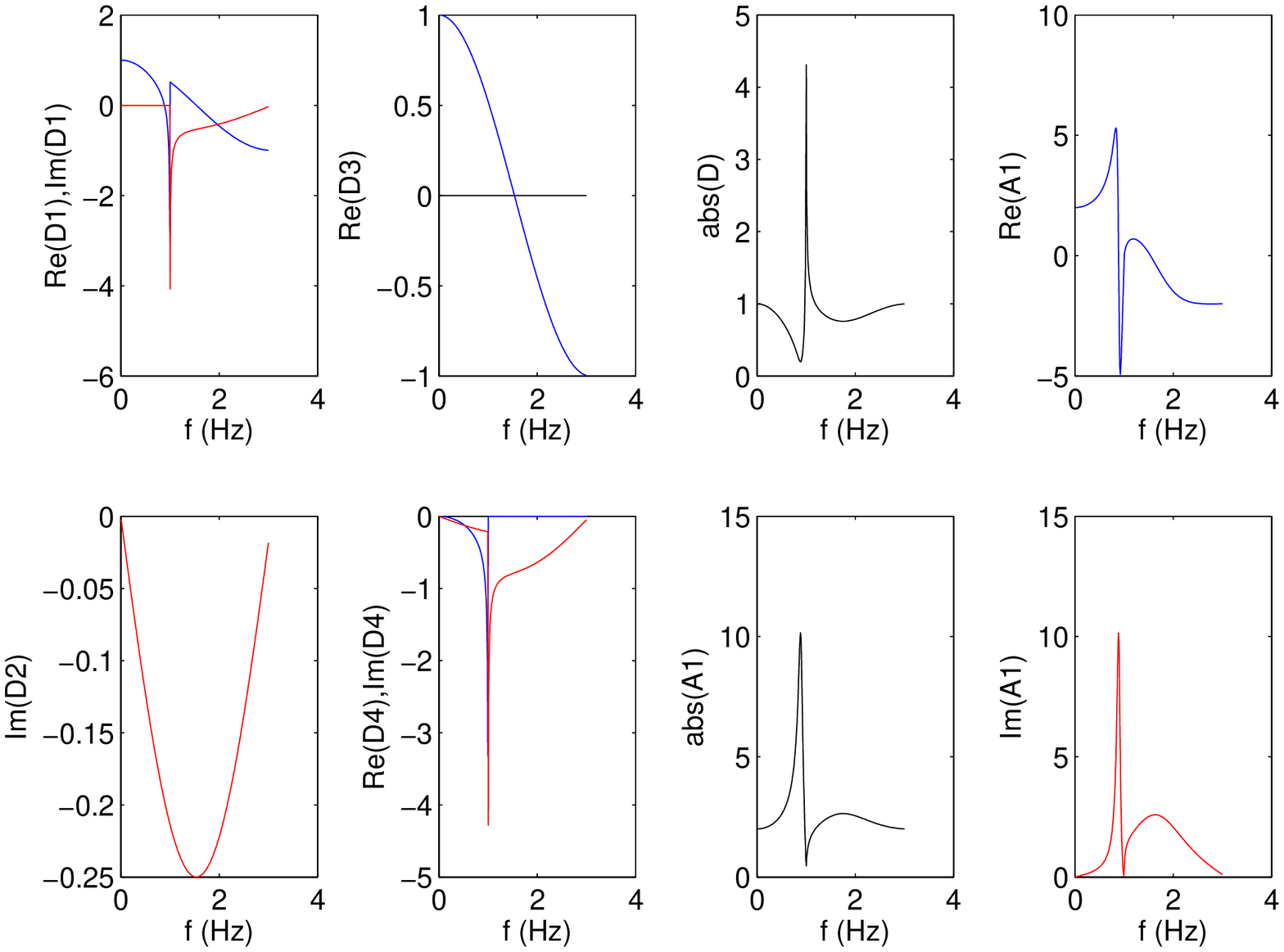}
\caption{The upper left-most panel depicts $\Re(D_{1}(f))$ in blue and  $\Im(D_{1}(f))$ in red. The second upper panel depicts $\Re(D_{3}(f))$ in blue and the zero level in black. The next upper panel depicts $\|D(f)\|$ and the right-most upper panel (an expanded version of which is given in the upper right-hand panel of the preceding figure) depicts $\Re(a_{0}^{[1](1)}(f))=\Re(A^{[1]}(f))$. The lower left-most panel depicts $\Im(D_{2}(f))$.  The second lower  panel depicts $\Re(D_{4}(f))$ in blue and  $\Im(D_{4}(f))$ in red. The next lower panel depicts $\|a_{0}^{[1](1)}(f)\|=\|A^{[1]}(f)\|$.  The right-most lower panel (an expanded version of which is given in the lower right-hand panel of the preceding figure) depicts $\Im(a_{0}^{[1](1)}(f))=\Im(A^{[1]}(f))$.  Case $w=430~m$,  $h=280~m$,}
\label{casest-08}
\end{center}
\end{figure}

\begin{figure}[ptb]
\begin{center}
\includegraphics[width=0.75\textwidth]{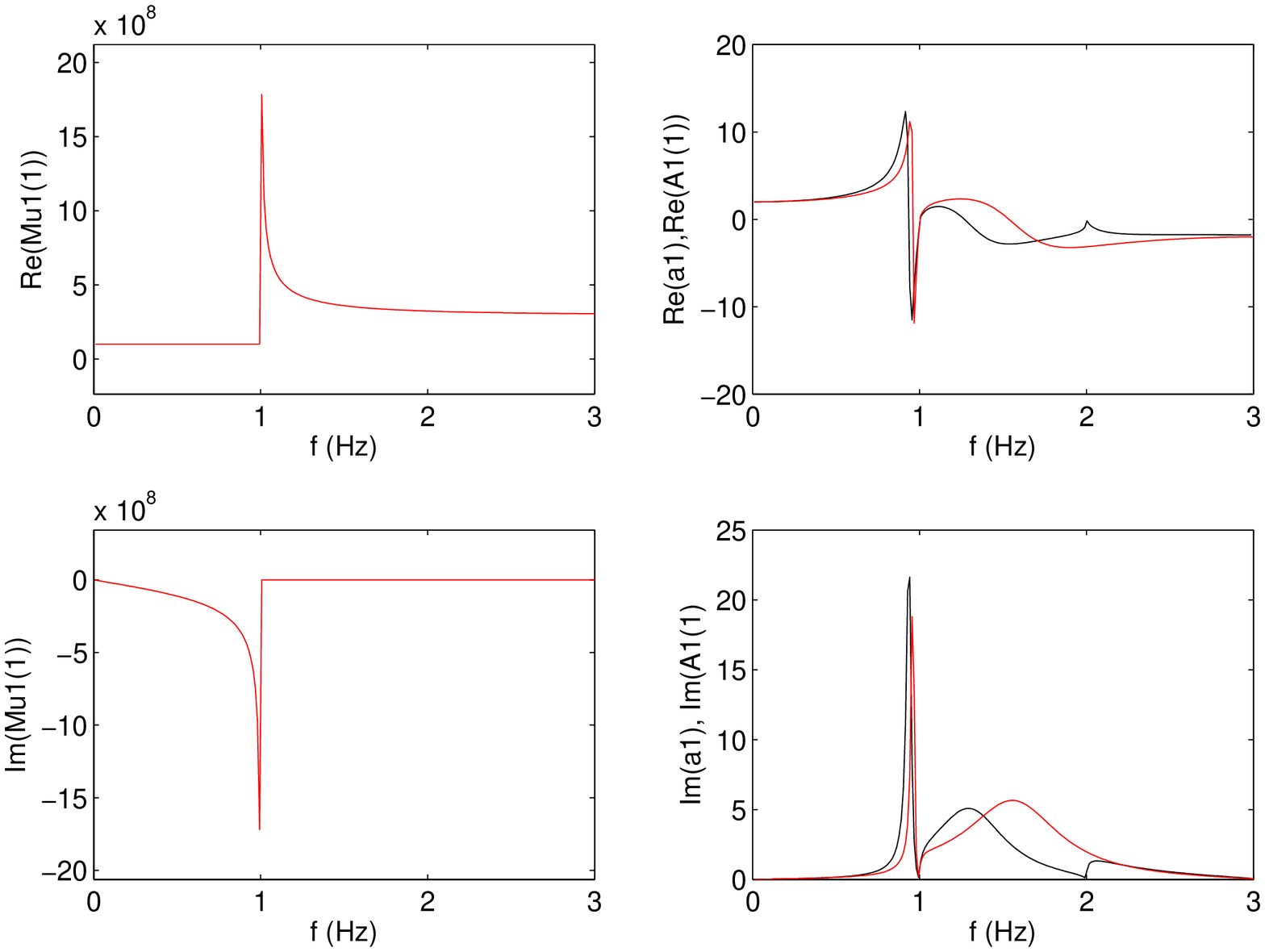}
\caption{Same as fig. \ref{casest-07} except that $w=172~m$,  $h=280~m$,}
\label{casest-09}
\end{center}
\end{figure}

\begin{figure}[ptb]
\begin{center}
\includegraphics[width=0.75\textwidth]{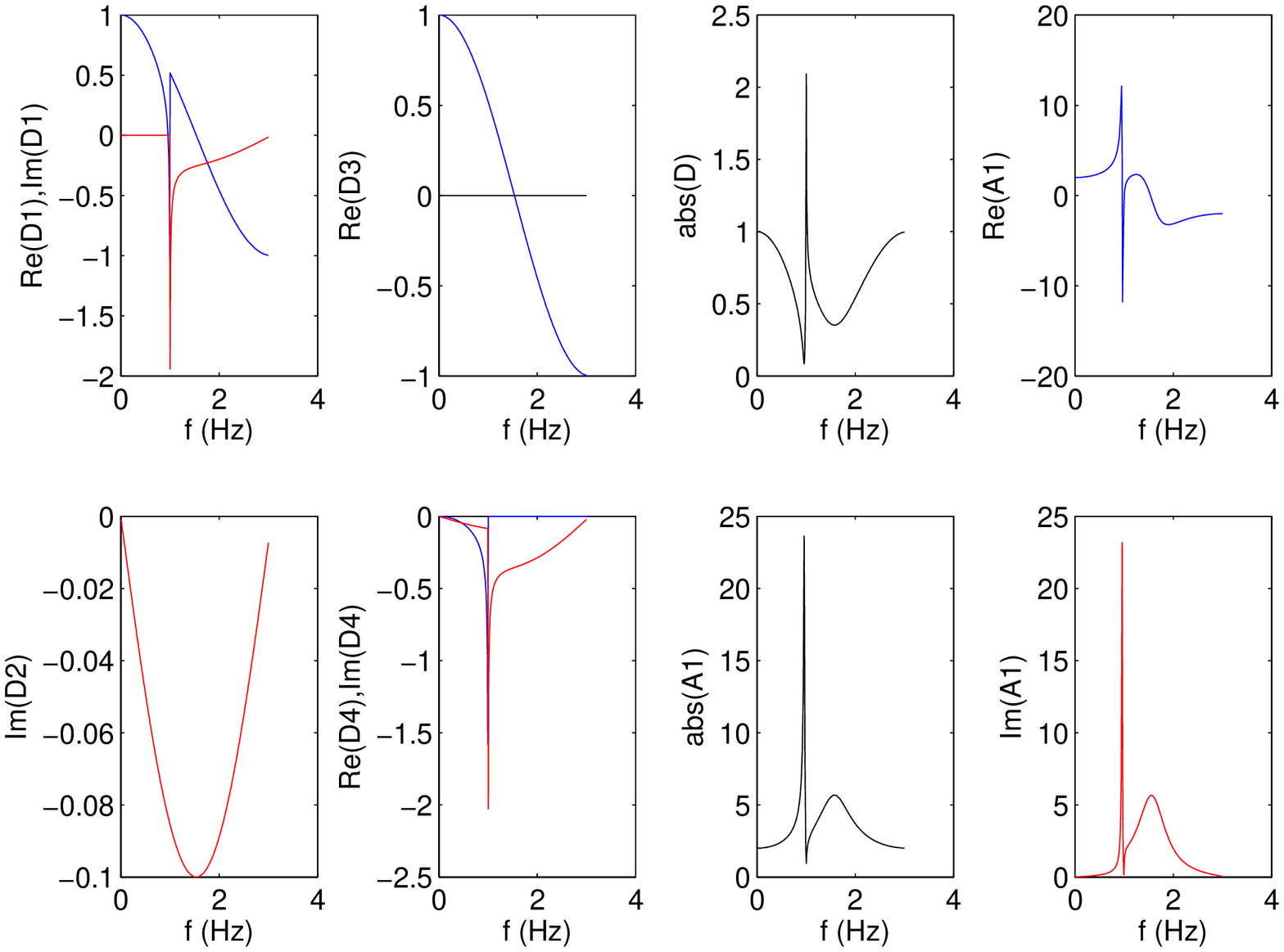}
\caption{Same as fig. \ref{casest-08} except that $w=172~m$,  $h=280~m$,}
\label{casest-10}
\end{center}
\end{figure}

\begin{figure}[ht]
\begin{center}
\includegraphics[width=0.75\textwidth]{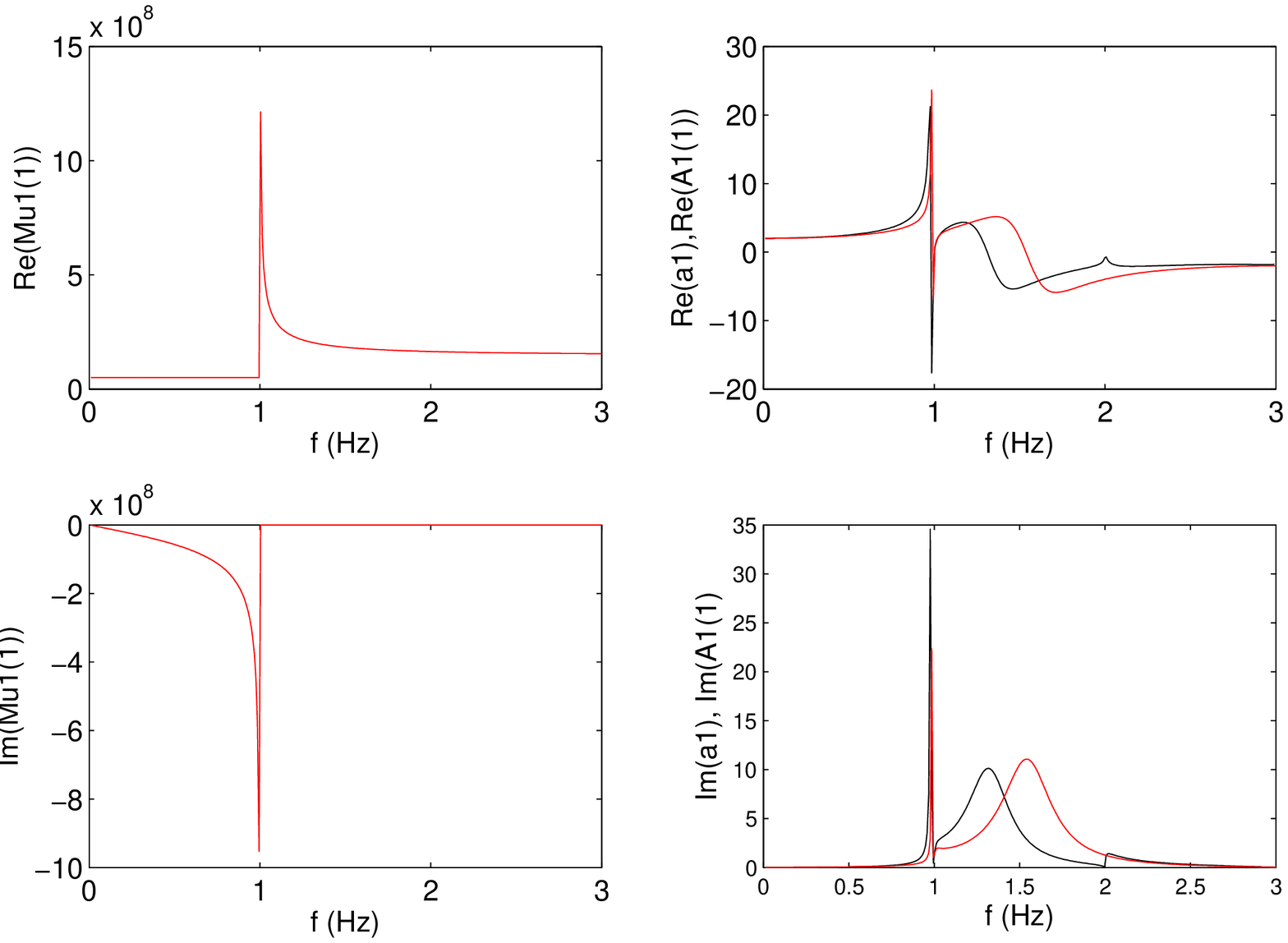}
\caption{Same as fig. \ref{casest-07} except that $w=86~m$,  $h=280~m$.}
\label{casest-11}
\end{center}
\end{figure}

\begin{figure}[ptb]
\begin{center}
\includegraphics[width=0.75\textwidth]{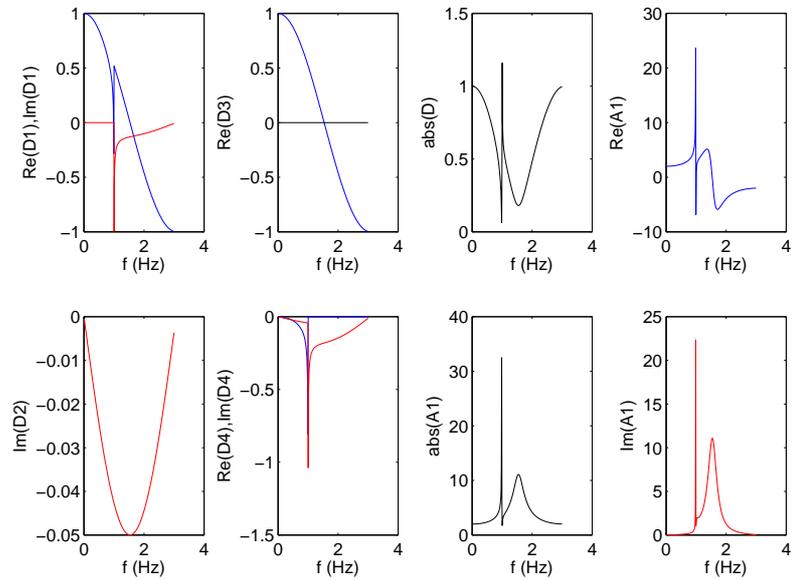}
\caption{Same as fig. \ref{casest-08} except that $w=86~m$,  $h=280~m$,}
\label{casest-12}
\end{center}
\end{figure}

Figs. \ref{casest-07}-\ref{casest-12} are relative to case studies of variation of $w$ for small-$h$ unveven boundaries. In all these figures, $\theta^{i}=0^{\circ}$, $d=1720~m$, $\beta^{[0]}=17200~ms^{-1}$, $\mu^{[0]}=1\times 10^{9}~Pa$,  $\beta^{[1]}=1720~ms^{-1}$, $\mu^{[1]}=1\times 10^{9}~Pa$ and the reference solutions  are obtained for $M=N=25$. We have omitted the arrows in these figures because the occurrences and locations of the LM and FBSWM resonances should, by now, be obvious.
\clearpage
\newpage
Consider the prototypical response in the low-frequency interval $[0~Hz,1.5~Hz]$ depicted in figs. \ref{casest-09}-\ref{casest-10}.  The modulus of this response is again characterized by two peaks, one, larger and narrower, to the left of, but very close to $f_{WF}=1~Hz$ due to the smaller (than previously) value of $h$, and the other, smaller and wider, to the right of, and farther from, $f_{WF}$. The left-hand peak is clearly due to the excitation of a LM whereas  the right-hand peak is obviously due to the excitation of a FBSWM . The imaginary parts of $D$ at these locations (smaller for the LM than for the FBSWM) explain why the LM peak is higher than the FBSWM peak.

This set of figures also shows that the narrower  is the uneven feature (i.e., the smaller is $w$) of the boundary, the higher and narrower are both  the LM  and FBSWM peaks, although the LM peaks are always higher and narrower than the FBSWM peaks, this being probably due to the fact that the LM peak indicates the occurrence of a true resonance whereas the FBSWM  peak that of a pseudo-resonance.

Finally, by comparing fig. \ref{casest-04} (for $h=560~m$) to fig. \ref{casest-08} (for $h=280~m,$), both relative to $w=430~m$, we see that The LM resonance peak of the latter is higher than that of the former which shows that the LM resonances are generally stronger for smaller $h$ ($w$ and $d$ being constant).
\section{Comments on the  the suitability of the effective (surrogate) configuration in the guise of a conclusion}
 Our rather simple theoretical analysis, has enabled the obtention of two significant results:\\
  (i) the explanation of how and why the $M=1$ approximation, contrary to the $M=0$ approximation, accounts for the lowest-frequency LM resonance and FBSWM pseudo-resonance, and \\ (ii) a simple, closed-form expression for the  shear modulus of the effective layer (in addition to the rather reasonable assumptions $1a-1e$). We now offer some comments on the nature of the effective configuration.\\\\
(a) Although the constitutive parameters $\beta,\mu$ of the uneven boundary configuration were assumed not to depend on $\omega$, the effective shear modulus, but not the effective body wave velocity, in the layer component of the surrogate configuration was found to depend on $\omega$. Such a dichotomy of effective parameters has been previously obtained in several publications amongst which  \cite{fb05}. The essential difference of our results with those of Felbacq and Bouchitt\'e  is that we are able to account for all the low-frequency resonances whereas their effective parameters fail (in their first numerical example) to account for one of the low-frequency resonances.\\\\
(b) The effective constitutive parameters $M^{[1]},~B^{[1]}$ do not depend on the incident angle (assumed to be identical in both the uneven boundary configuration and the surrogate configuration) because our analysis, like those of many others relying on the NRW technique, was based on the supposition $\theta^{i}=\Theta^{i}=0$. The reason for this supposition is that it enables an explicit, closed-form expression to be obtained for $M^{[1]}$ which is less-easy to find for non-normal incidence.\\\\
(c) We obtained $M^{[1]},~B^{[1]}$ via the $M=0$ and $M=1$ approximations of the uneven boundary response and did not carry out the analysis for larger $M$ because our aim was essentially to show that  one resonance and one pseudo-resonance could be accounted for by this approximation procedure (and by the effective parameters to which it leads). In any case, the numerical results in the first part of this study show clearly that the $M=2$ approximation enables to account for two LM resonances, etc., but we made no attempt, in this study, to show theoretically why this is so.\\\\
(d) Although the shear modulus $\mu$ of the uneven boundary configuration was assumed to be real, the effective shear modulus $M^{[1]}$ is generally- complex; in fact it is complex for the low frequencies corresponding to $k<2\pi/d$ and real for the higher frequencies corresponding to $k>2\pi/d$ as shown graphically in figs. \ref{numMeq1-01}-\ref{numMeq1-03}.\\\\
(e) Since sinc$(\pi)=0$, the $M^{[1]}$ obtained via the $M=1$ approximation tends towards the $M^{[1]}$ obtained via the $M=0$ approximation as $w/d\rightarrow 1$.\\\\
(f) Since $\lim_{f\rightarrow 0}\frac{k}{k_{z1}^{[1]}}=0$, the the $M^{[1]}$ obtained via the $M=1$ approximation tends towards the $M^{[1]}$ obtained via the $M=0$ approximation as $f\rightarrow 0$.\\\\
(g) Both $M^{[1]}$ and $B^{[1]}$ do not depend on the layer thickness=amplitude of the boundary uneveness, which is a favorable result in that one usually wants an effective medium to not depend on its thickness (otherwise, how qualify it as being a 'medium'?).\\\\
(h) As seems reasonable, all other parameters (i.e., those related to the solicitation and to the medium in $\Omega_{0}$) of the surrogate configuration are equal to those of the uneven boundary configuration, which means, amongst other things, that the essential features of the  effective medium are those of the surrogate layer, which, although being homogeneous, is generally lossy and dispersive, this lossy, dispersive nature compensating to a large extent for the inhomogeneous nature (i.e., uneveness of the region located between $z=0$ and $z=h$) of the uneven boundary configuration.\\\\
(i) The gain, from the point of view of understanding the origin of the complex response phenomena produced by an uneven boundary, is substantial since predicting the response of a configuration comprising a homogeneous layer over a homogeneous half space is simple, and mathematically explicit at all frequencies, whereas predicting the response of the uneven boundary configuration is complicated and mathematically unexplicit (i.e., requires solving numerically a large system of coupled linear equations).\\\\
(j) The notion of effective medium is often tied up with the identity of the fields in the surrogate and original configurations. This is, of course, impossible throughout the domains included between the two boundaries $z=0$ and $z=h$ as underlined in (\ref{11-260}) by the fact that the equality holds only in the intervals $nd-\frac{w}{2}\le x\le nd+\frac{w}{2}~,n=0,\pm 1,....$. Naturally, outside these $x-$ intervals wherein the medium is the vacuum, the elastic wavefield is nil whereas in the layer configuration it does not vanish outside of  $nd-\frac{w}{2}\le x\le nd+\frac{w}{2}~,n=0,\pm 1,....$. What this means is that our homogenization procedure does not assume equality of two elastic field between $z=0$ and $z=h$ for all $x$, but rather that of two observables: the fields at a specific point on $z=h$ of the uneven boundary configuration and at the same point of the layer configuration. These locations are in the near-field region of the scattering configuration whereas the observables employed in the NRW technique are in the far-field zone. We could have proceeded as in the NRW technique by comparing the far-field observables $a_{0}^{[0]-}$ to $A^{[0]-}$ but chose not to do so because in typical geophysical and other elastic wave problems the observables are of the near-field variety (e.g., in terrestrial near-surface (e.g., petro-) geophysics,  the sensors are typically at or near the earth's surface and the phenomena that are measured are often due to heterogeneities near the surface).\\\\
(k) the dispersion of the effective shear modulus is not anomalous since the  real part of the effective shear modulus is positive and greater or equal to $\mu\frac{w}{d}$ for all (low) frequencies; by the same means one shows that the imaginary part of the effective shear modulus is negative or nil for these frequencies. This is at odds with the anomalous dispersion found in an effective constitutive parameter in such publications as \cite{fb05}.\\\\
(l) Despite its rather simple nature, the effective shear modulus (of the layer above the solid half space) obtained from the $M=1$ approximation of near-field response of the uneven boundary is capable (contrary to the effective shear modulus obtained from the $M=0$ approximation) of accounting quite accurately for the two essential peak-like features (in terms of location, height and width of the peaks) that condition the said response at low frequencies: the Love mode resonance and the Fixed-base shear wall pseudoresonance, this being so for a wide range of boundary uneveness parameters such as $h$ and $w/d$.\\\\
(m) The analysis presented herein may shed some useful light on the  problems of seismic response in such geophysical configurations as hill and mountain ranges, cities, etc.

\end{document}